\documentclass[12pt]{iopart}
\usepackage{iopams,graphicx}
\usepackage{mathcomp} %% for non-italic unit 'microliter' : $\tcmu$l

\begin{document}

\title{Rolling ferrofluid drop on the surface of a liquid}
\author{V Sterr$^1$, R Krau{\ss}$^2$,  K I Morozov$^3$, I Rehberg$^2$, A Engel$^1$,
        R Richter$^2$,}
\address{$^1$ Institut f\"ur Physik, Carl von Ossietzky Universtit\"at, 26111
        Oldenburg, Germany}
\address{$^2$ Experimentalphysik V, Universit\"at Bayreuth, 95440 Bayreuth, Germany}
\address{$^3$ Institute of Continuous Media Mechanics, 1 Korolev Street, 614013 Perm,
         Russia}
\ead{\mailto{sterr@theorie.physik.uni-oldenburg.de},\\
\mailto{reinhard.richter@uni-bayreuth.de}}
\date{\today}

%%%%%%%%%%%%%%%%%%%%%%%%%%%%%%%%%%%%%%%%%%%%%%%%%%%%%%%%%%%%%%%%%%%%%%%%%%%%%%%%%%%%%%
%%%%%%%%%%%%%%%%%%%% NEWCOMMANDS %%%%%%%%%%%%%%%%%%%%%%%%%%%%%%%%%%%%%%%%%%%%%%%%%%%%%

%% certain vectors
\newcommand\vecom{\bi{\Omega}}
\newcommand\vv{\bi{v}}
\newcommand\vecr{\bi{r}}
\newcommand\vA{\bi{A}}
\newcommand\vAlm{\bi{A}_{\ell m}}
\newcommand\vAtp{\bi{A}_{\ell m}(\vartheta,\varphi)}
\newcommand\vAlmp{\bi{A}_{\ell' m'}}
\newcommand\vB{\bi{B}_{\ell m}}
\newcommand\vBlm{\bi{B}_{\ell m}}
\newcommand\vBtp{\bi{B}_{\ell m}(\vartheta,\varphi)}
\newcommand\vBlmp{\bi{B}_{\ell' m'}}

%% velocity components
\newcommand\vr{v_r}
\newcommand\vth{v_\vartheta}
\newcommand\vphi{v_\varphi}
\newcommand\vpsi{v_\psi}
\newcommand\vz{v_z}

%% misc.
\newcommand\itom{{\it\Omega}}
\newcommand\ehoch{\mathrm{e}^}
\newcommand\ii{\rmi}
\newcommand\of{(r,\vartheta,\varphi)}
\newcommand\mysum{\sum_{\ell=0}^{\infty}\sum_{m=-\ell}^{+\ell}}
\newcommand\dlm{\frac{\rmd^{\ell+m}\phantom{(\cos\vartheta)}}{\rmd(\cos\vartheta)^{\ell+m}}}
\newcommand\la{\langle}
\newcommand\ra{\rangle}
\newcommand\einsop{1\hspace{-0.4em}1}
\newcommand\Nat{N{\hspace{-1.1em}I\;}}
\newcommand\Reell{R{\hspace{-0.9em}I\quad}}
\newcommand\transp{^\textrm{\footnotesize T}}
\newcommand\ls{L_\mathrm{s}}
\newcommand\imag{\mathfrak{Im}}
\newcommand\real{\mathfrak{Re}}

%% equations
\newcommand\bq{\begin{equation}}
\newcommand\eq{\end{equation}}
\newcommand\bqa{\begin{eqnarray}}
\newcommand\eqa{\end{eqnarray}}

%% differentials
\newcommand\dt{\rmd t}
\newcommand\du{\rmd u}
\newcommand\dx{\rmd x}
\newcommand\dth{\rmd\vartheta}
\newcommand\dphi{\rmd\varphi}
\newcommand\de{\rmd}

%% partials
\newcommand\pa{\partial}
\newcommand\parr{\,\partial_r}
\newcommand\parth{\,\partial_\vartheta}
\newcommand\parphi{\,\partial_\varphi}
\newcommand\parpsi{\,\partial_\psi}

%% Einheitsvektoren
\newcommand\ei{\mathbf{{e}}_i}
\newcommand\ej{\mathbf{{e}}_j}
\newcommand\eix{\mathbf{{e}}_x}
\newcommand\ey{\mathbf{{e}}_y}
\newcommand\ez{\mathbf{{e}}_z}
\newcommand\er{\mathbf{{e}}_r}
\newcommand\etheta{\mathbf{{e}}_\vartheta}
\newcommand\ephi{\mathbf{{e}}_\varphi}
\newcommand\epsi{\mathbf{{e}}_\psi}

%% radial functions
\newcommand\flm{f_{\ell m}}
\newcommand\glm{g_{\ell m}}
\newcommand\hlm{h_{\ell m}}
\newcommand\iflm{f_{\ell m}^{(\mathrm{i})}}
\newcommand\iglm{g_{\ell m}^{(\mathrm{i})}}
\newcommand\ihlm{h_{\ell m}^{(\mathrm{i})}}
\newcommand\oflm{f_{\ell m}^{(\mathrm{o})}}
\newcommand\oglm{g_{\ell m}^{(\mathrm{o})}}
\newcommand\ohlm{h_{\ell m}^{(\mathrm{o})}}
\newcommand\flmmi{f_{\ell,- m}}
\newcommand\glmmi{g_{\ell,- m}}
\newcommand\hlmmi{h_{\ell,- m}}
\newcommand\iflmmi{f_{\ell,- m}^{(\mathrm{i})}}
\newcommand\iglmmi{g_{\ell,- m}^{(\mathrm{i})}}
\newcommand\ihlmmi{h_{\ell,- m}^{(\mathrm{i})}}
\newcommand\oflmmi{f_{\ell,- m}^{(\mathrm{o})}}
\newcommand\oglmmi{g_{\ell,- m}^{(\mathrm{o})}}
\newcommand\ohlmmi{h_{\ell,- m}^{(\mathrm{o})}}

%% coefficients
\newcommand\alm{a_{\ell m}}
\newcommand\blm{b_{\ell m}}
\newcommand\clm{c_{\ell m}}
\newcommand\ddlm{d_{\ell m}}
\newcommand\pplm{p_{\ell m}}
\newcommand\qlm{q_{\ell m}}
\newcommand\Blm{B_{\ell m}}

\newcommand\almmi{a_{\ell, -m}}
\newcommand\blmmi{b_{\ell, -m}}
\newcommand\clmmi{c_{\ell, -m}}
\newcommand\ddlmmi{d_{\ell, -m}}
\newcommand\pplmmi{p_{\ell, -m}}
\newcommand\qlmmi{q_{\ell, -m}}
\newcommand\Blmmi{B_{\ell, -m}}

%% spherical harmonics
\newcommand\ytp{Y_{\ell m}(\vartheta,\varphi)}
\newcommand\ytpmi{Y_{\ell,-m}(\vartheta,\varphi)}
\newcommand\ytpst{Y_{\ell m}^*(\vartheta,\varphi)}
\newcommand\yststr{Y_{\ell' m'}^*(\vartheta,\varphi)}
\newcommand\ylm{Y_{\ell m}}
\newcommand\ylmp{Y_{\ell' m'}}
\newcommand\ylmmi{Y_{\ell,-m}}

%% Legendre functions
\newcommand\pth{P_{\ell m}(\cos\vartheta)}
\newcommand\pthplus{P_{\ell, m+1}(\cos\vartheta)}
\newcommand\pthmi{P_{\ell, m-1}(\cos\vartheta)}
\newcommand\plm{P_{\ell m}}
\newcommand\pleins{P_{\ell 1}}
\newcommand\plnull{P_{\ell 0}}
\newcommand\klm{K_{\ell m}}
\newcommand\kleins{K_{\ell 1}}

%% sine, cosine
\newcommand\sth{\sin\vartheta}
\newcommand\sqth{\sin^2\vartheta}
\newcommand\sph{\sin\varphi}
\newcommand\smph{\sin(m\varphi)}
\newcommand\sps{\sin\psi}
\newcommand\smps{\sin(m\psi)}
\newcommand\cth{\cos\vartheta}
\newcommand\cph{\cos\varphi}
\newcommand\cmph{\cos(m\varphi)}

%% stress tensor components
\newcommand\sigrr{\sigma_{rr}}
\newcommand\sigrt{\sigma_{r\vartheta}}
\newcommand\sigrp{\sigma_{r\varphi}}
\newcommand\sigrps{\sigma_{r\psi}}
\newcommand\sigpt{\sigma_{\varphi\vartheta}}

%%%%%%%%%%%%%%%%%%%%%%%%%%%%%%%%%%%%%%%%%%%%%%%%%%%%%%%%%%%%%%%%%%%%%%%%%%%%%%%%%%%%%%%
%%%%%%%%%%%%%%%%%%%%%%%%%%%%%%%%%%%%%%%%%%%%%%%%%%%%%%%%%%%%%%%%%%%%%%%%%%%%%%%%%%%%%%%

\begin{abstract}

We report on the controlled transport of drops of magnetic liquid, which are
swimming on top of a non-magnetic liquid layer. A magnetic field which is
rotating in a vertical plane creates a torque on the drop. Due to
surface stresses within the immiscible liquid beneath, the drop is propelled
forward. We measure the drop speed for different field amplitudes, field
frequencies and drop volumes. Simplifying theoretical models describe the drop
either as a solid sphere with a Navier slip boundary condition, or as a liquid
half-sphere. An analytical expression for the drop speed is obtained which is
free of any fitting parameters and is well in accordance with the experimental
measurements. Possible microfluidic applications of the rolling drop are also
discussed.
\end{abstract}


\submitto{\NJP}

%\maketitle

%%%%%%%%%%%%%%%%%%%%%%%%%%%%%%%%%%%%%%%%%%%%%%%%%%%%%%%%%%%%%%%%%%%%%%%%%%%%%%%%%%%%%%%%%%%%%%%%%%%%%%%%%%%%%%%%%%%%%%%%%%%%%%%%%%%%%%%%%%%
%%%%%%%%%%%%%%%%%%%%%%%%%%%%%%%%%%%%%%%%%%%%%%%%%%%%%%%%%%%%%%%%%%%%%%%%%%%%%%%%%%%%%%%%%%%%%%%%%%%%%%%%%%%%%%%%%%%%%%%%%%%%%%%%%%%%%%%%%%%

\section{Introduction} \label{sec_intro}
A tiny drop of magnetic fluid responds to magnetic fields in many ways -- it is
a "world in a nutshell". Typically 1\,$\tcmu$l of magnetic fluid (MF) contains
more than $10^{13}$ magnetic mono-domain particles, each with a diameter of around
$10$\,nm, which are suspended in a carrier fluid like water or kerosene
\cite{rosensweig1985}.
% \Phi \approx 0.01 and r_particle \approx 5 nm  --> N \approx 0.2 10^14
In the absence of an external magnetic field there is no long-range order in
the MF, but when exposed to a static field the magnetic grains orient in part which
results in a net magnetization.
%\cite{blums1997} ???
Application of a rotating magnetic field induces a torque on the suspended
magnetic grains. Due to the viscous coupling of the particles to its
surrounding carrier liquid angular momentum is transferred to the whole drop
and an abundance of phenomena is observed.

In a series of experiments pioneered by Bacri \emph{et al}~\cite{bacri1994} a
magnetic drop was levitated in a surrounding liquid and exposed to a field
rotating in the horizontal plane. For very small angular frequencies of the field
an elongated drop follows the field rotation quasi-adiabatically with small
phase lag
\cite{cebers1995,lebedev1997,morozov1997,sandre1999,cebers2002}. % citations 13-15 nachtragen!
In the limit of high angular frequency one observes for small magnetic fields
an oblate spheroid, for intermediate values transient shapes, and for large
fields an oblate spheroid with "spiny starfish" appearance
\cite{bacri1994,morozov2000,lebedev2003}, for a review see
Ref.\,\cite{blums1997}.

%Recently we have shown experimentally that a field which is rotating in a
%\emph{vertical plane} can drive a flow of ferrofluid in a horizontal duct by
%means of the magnetic torque wich generates a surface stress
%\cite{krauss2005,krauss2006}. The experiment performed and analyzed in the
%following article is derived from this configuration: A drop of

Our setup, investigated in experiment and theory in this article, differs from
the above configuration in two points fundamentally: (i) the field is rotating
in a plane \emph{oriented vertically}, (ii) the drop of ferrofluid is swimming
\emph{on top} of a layer of non-magnetic fluid. The field configuration is
borrowed from a recent experiment ("the magnetic pump") where the magnetic
torque drives a continuous flow of ferrofluid in an open duct
\cite{krauss2005,krauss2006}. By replacing the ferrofluidic layer with a
floating drop we are able to propel the drop with a constant translation
velocity $\vv_\mathrm{drop}$ with respect to the liquid surface. Moreover we
could in principle manoeuvre the drop to arbitrary positions on the whole two
dimensional liquid layer by utilizing an additional alternating field in
y-direction. This is a new and promising technique for microfluidic
applications.

Our theoretical model describes the ferrofluid drop first as a solid
sphere with a Navier slip boundary condition at its surface, then as a liquid
(half-)sphere with own inner flow field. The problem is treated within Stokes
approximation and the assumption of certain symmetries. In both cases
an analytical expression for the drop speed $\vv_\mathrm{drop}$ in terms
of the experimentally accessible parameters is obtained. While the solution
of the Navier slip model contains an unknown parameter, the slip length,
the result of the liquid half-sphere model is completely free of fitting parameters
and is shown to represent the experimentally measured dependencies very well.

\begin{figure}[tb]
\centering
\includegraphics[width=11cm]{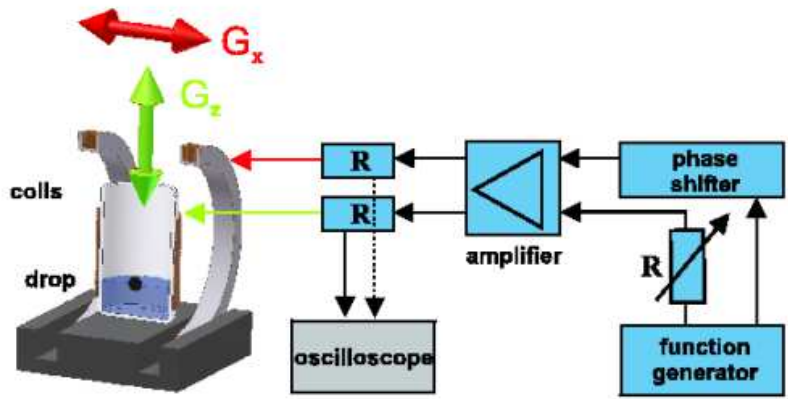}
\caption{Sketch of the experimental setup. For details see text.}
\label{setup}
\end{figure}

The article is organized as follows. Next we present the experimental
arrangement together with some qualitative observations. This is followed by a
comprehensive theoretical analysis (section~\ref{sec_theory}). Subsequently the
results obtained by experiment and theory are compared in section~\ref{sec_comp}
and discussed in section~\ref{sec_conc}.
%%%%%%%%%%%%%%%%%%%%%%%%%%%%%%%%%%%%%%%%%%%%%%%%%%%%%%%%%%%%%%%%%%%%%%%%%%%%%%%%%%%%%%%%%
%%%%%%%%%%%%%%%%%%%%%%%%%%%%%%%%%%%%%%%%%%%%%%%%%%%%%%%%%%%%%%%%%%%%%%%%%%%%%%%%%%%%%%%%%

\section{Experiment}\label{sec_experiment}

Our experimental setup is shown in figure~\ref{setup}. We place a cylindrical
glass beaker in between a Helmholtz pair of coils that produce an externally
applied field $G_x(t)$ which is oriented horizontally. In addition another coil
is wrapped directly around the beaker providing a field $G_z(t)$ in vertical
direction. Here we denote the external magnetic far field by $\bi{G}$ and the
local one by $\bi{H}$. A sinusoidal driving current is supplied by connecting
the output of a function generator (Fluke PM 5138A) to one channel of a power
amplifier (Rotel RB-1090). The input of the second channel is supplied with a
delayed signal of the function generator. In order to allow an independent
adjustment of both currents, a variable resistor is inserted in one driving
circuit. An oscilloscope serves to control the phase difference of both
currents. When the phase difference is set to 90$^o$ the coils produce a
rotating field $\bi{G}(t)$ inside the beaker. Any motion of the drop of
magnetic liquid is observed from above by means of a video camera (not shown
here).

For a good performance of the driving by the rotating field a large imaginary
part of the susceptibility of the MF is important. Thus we have selected a
magnetic fluid based on air stable cobalt particles \cite{boennemann2003b},
which are stabilized by oleic acid in kerosene. Figure~\ref{fig:chi} reproduces
the frequency dependence of the complex susceptibility of this fluid measured
by an ac-susceptometer \cite{krauss2006}. The MF has a volume fraction of
5\,$\%$ and constitutes the interior ${\mathrm{(i)}}$ of the drop. Its
viscosity was determined to be $\eta^\mathrm{(i)}=5.4$\,mPa\,s by means of a
low shear rheometer (Contraves LS40), and the density of the MF has been found
to be $\rho^\mathrm{(i)}=1.07$\,gcm$^{-3}$.

\begin{figure}[tb]
\begin{minipage}{0.45\textwidth}
\includegraphics[width=10cm]{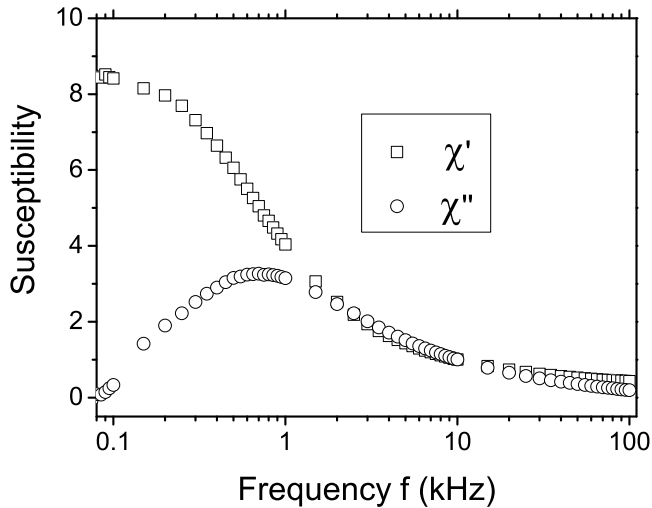}
\end{minipage}\hspace{1cm}
\begin{minipage}{0.45\textwidth}
\begin{flushleft}
\caption{The magnetic susceptibility of the cobalt based magnetic liquid versus
the external alternating magnetic field. The data points for the real and
imaginary parts of the susceptibility are marked by squares and circles,
respectively.} \label{fig:chi}
\end{flushleft}
\end{minipage}
\end{figure}

The drop of MF has to float on top of a liquid layer of a non-magnetic fluid.
The quantities of this fluid outside of the drop will be marked by $\mathrm{(o)}$.
This fluid must not mix with any of the components of the MF.
Moreover it must be denser than the MF. A per-fluorinated hydrocarbon fluid
(Galden SV-90) proved as suitable substrate because of its higher density
$\rho^\mathrm{(o)}=1.69$\,gcm$^{-3}$, its long-term stability, and its
non-miscibility with the MF. According to the manufacturer the viscosity
amounts to $\eta^\mathrm{(o)}=1.27$\,mPa\,s, and the surface tension to
$\gamma=16$\,mN/m. This fluid is poured into a cylindrical glass beaker up to a height
of 2\,cm in order to minimize fringe effects from the bottom of the glass.

At the beginning of an experiment a definite volume $V$ of MF is put on the
surface of the per-fluorinated liquid with a pipette. According to the density
ratio of the two liquids the forming drop immerses with approximately two
thirds of its volume (corresponding to a measured maximum penetration depth of
about 60\,\% of its diameter). The rotating field generated by the coils leads
to a motion of the droplets in the direction the field is rolling. Hence the
direction of the motion can be reversed by changing the sign of the phase
difference between the ac-fields. Under the given experimental conditions we
can achieve droplet velocities up to a few cm/s. The good contrast between the
black MF and the transparent hydrocarbon liquid allows an easy observation by a
digital video camera. Two exemplary movies can be activated at
figure~\ref{fig:movie}. The velocity of the droplets was determined by
extracting the time a drop takes to travel the distance of 1\,cm in the center
of the beaker. Within this distance the magnetic field varies less than 1\,\%.

\begin{figure}[tb]
\centering
(a)
\includegraphics[width=0.4\textwidth]{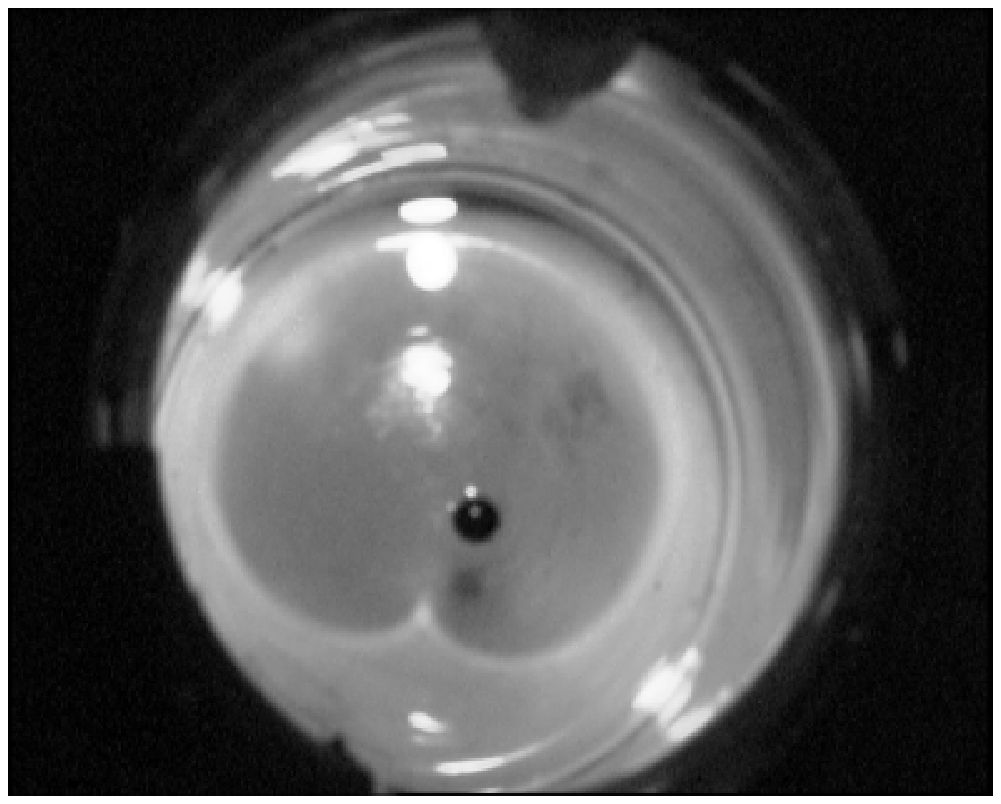}
(b)
\includegraphics[width=0.4\textwidth]{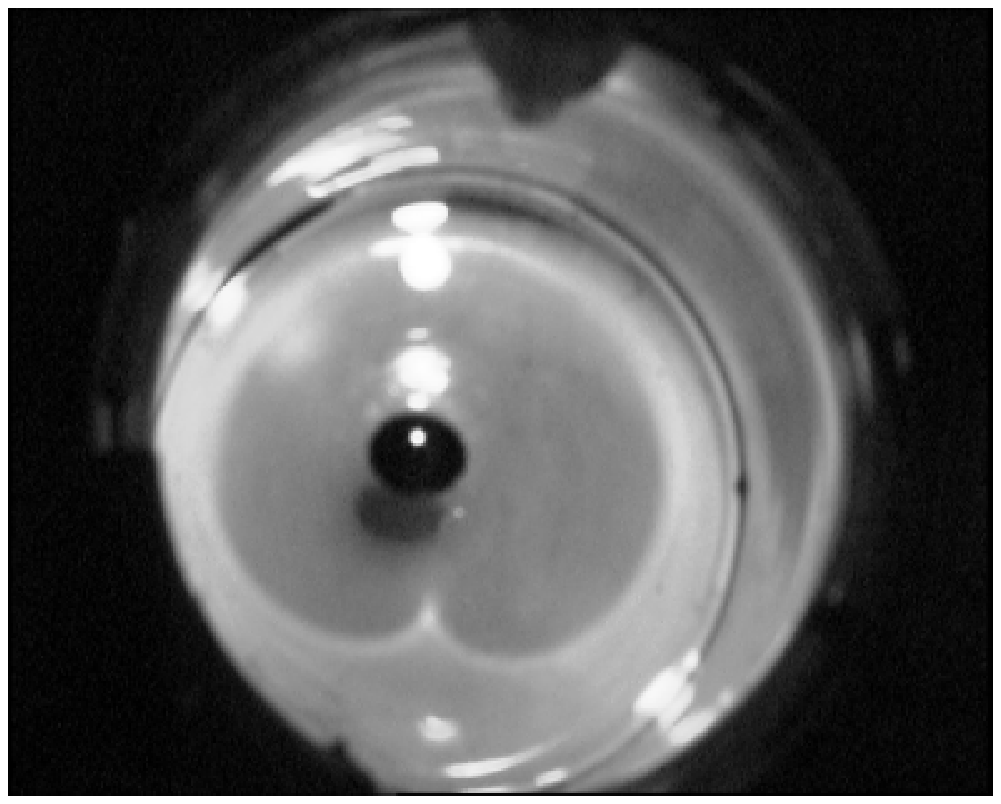}
\caption{Drops of magnetic fluid with a volume of (\textbf{a}) $5\,\tcmu$l
(see \underline{movie1}) and (\textbf{b}) $80\,\tcmu$l (\underline{movie2})
are rolling on top of a per-fluorinated Newtonian liquid.} \label{fig:movie}
\end{figure}
%%%%%%%%%%%%%%%%%%%%%%%%%%%%%%%%%%%%%%%%%%%%%%%%%%%%%%%%%%%%%%%%%%%%%%%%%%%%%%%%%%%%%%%
%%%%%%%%%%%%%%%%%%%%%%%%%%%%%%%%%%%%%%%%%%%%%%%%%%%%%%%%%%%%%%%%%%%%%%%%%%%%%%%%%%%%%%%

\section{Theory}\label{sec_theory}

The theoretical description of the "real" setup poses a very complicated
boundary value problem which would have to be solved by numerical methods. In
order to extract the essence of the effect we make some simplifying assumptions
which even lead to an analytical solution.

The droplet is considered to be a spherical object half-way immersed into a
liquid with an otherwise perfectly flat surface. Effects of gravity are
neglected as is the inertia term in the Navier-Stokes equation which is hence
rendered linear. This Stokes approximation is in order when the Reynolds number
$\mathrm{Re}$ is sufficiently small. Here it is given by $\mathrm{Re} =\itom
R^2\varrho^\mathrm{(o)}/\eta^\mathrm{(o)}$, with $\itom$ the angular velocity
of the sphere, and ranges between one and ten. The problem is treated within
the reference frame where the sphere is rotating with its center at rest
(cf.~figure\,\ref{sphere}). In order to ensure stationarity in this frame, the
overall forces and torques acting on the sphere must cancel out. After the
velocity field of the surrounding liquid has been determined, its asymptotic
value at $r\to\infty$ will give the negative translation velocity of the sphere
in the laboratory frame.
\begin{figure}[tb]
\begin{minipage}{0.45\textwidth}
  \includegraphics[width=10cm]{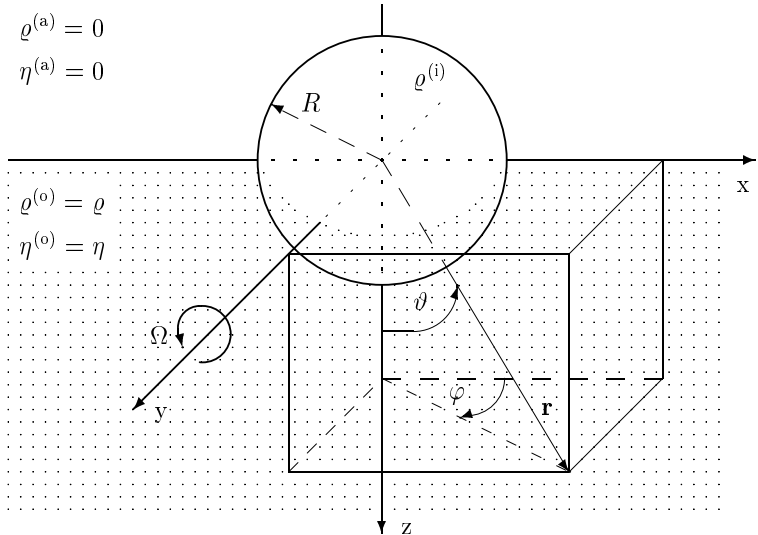}
\end{minipage}\hspace{1cm}
\begin{minipage}{0.45\textwidth}
\begin{flushleft}
\caption{A spherical ferrofluid drop with radius $R$ hosts in its inner (i) a
fluid with density $\varrho^\mathrm{(i)}$. It is covered from above (a) by a
gas with density $\varrho^\mathrm{(a)}$. The lower part of the drop is half-way
immersed into an outer (o) Newtonian fluid with density $\varrho^\mathrm{(o)}$
and dynamic viscosity $\eta^\mathrm{(o)}$. The drop rotates with constant
angular velocity $\itom$. The center of the sphere is the origin of the
reference frame as indicated in the picture.}
  \label{sphere}
\end{flushleft}
\end{minipage}
\end{figure}

The simplest approach is treating the droplet as a \emph{solid} sphere and employing
the common \emph{no-slip} boundary condition at its surface, but this would lead to a
logarithmically divergent viscous torque \cite{DA_sterr}.
It has long been shown \cite{huh_scri}, that
hydrodynamic problems containing a moving contact line in combination with the no-slip
condition give rise to diverging quantities due to an inherent contradiction:
on the one hand the fluid is supposed to stick to the solid surface,
and yet the line where solid, liquid, and gas meet shall advance on that very same
surface.

Several means have been proposed to relieve the singularity, e.g., taking into
account a strong curvature of the fluid surface near the solid, or describing
the contact region in terms of molecular interactions, as has been done in
Ref.\,\cite{deGen1}. A straightforward approach is to allow a certain amount of
slippage over the solid surface. As early as 1823, a linear relation between
the tangential stresses at the solid surface and the velocity of the latter was
proposed by C.-L.~Navier \cite{navier}. Although other forms of slip condition
can be successful \cite{dussan1,pis_rub} this "Navier slip" has become the most
popular one and has since been examined and applied oftentimes. Earlier works
distinguish between several regions where different expansions are made, and
only employ the slip condition in the contact region itself, finally matching
the solutions together \cite{huh_mas, hocking2,
  cox}. Our treatment, however, will follow the lines of Ref.\,\cite{oneill}
who applied the Navier slip condition on the whole solid surface without
seperating different regions. This is justified a posteriori by the fact that
the slippage shows most of its impact in the direct vicinity of the contact
line where the stresses are largest and leaves the flow field undisturbed
further away, as will be made clear by the results of the present paper.

Although Ref.\,\cite{oneill} considered a problem quite analogous to ours,
\emph{i.e.}, the rotation and translation of a solid spherical object which is
half-way immersed in a liquid, we will present the treatment in a more lucid
albeit less general way that will lead to a closed expression for the resulting
flow field which is lacking in Ref.\,\cite{oneill}.

The disadvantage of the Navier slip condition is that it contains a
characteristic length $\ls$ which is supposed to be small compared to the
length scales characterizing the problem (in our case the sphere radius $R$)
and essentially indicates how much the fluid molecules slip over the solid
surface. $\ls\rightarrow 0$ is equivalent to no slip, while
$\ls\rightarrow\infty$ corresponds to completely unimpeded slip or zero
tangential stress. This \emph{slip length} does not necessarily "represent true
slippage but merely recognizes the fact that the liquid consists of molecules
of finite size", as stated by Huh and Mason in Ref.\,\cite{huh_mas}. Or as Cox
puts it in Ref.\,\cite{cox}: "Slip between liquid and solid is a convenient
assumption to get rid of the non-integrable stress singularity." Although the
slip length between certain materials can be measured by now (see
e.g.~\cite{denn,jos_tab, schmatko,fetzer_etal}), this is of no use to the
present problem, as the experiment does not involve a solid sphere.

By consequence, the result of these calculations will not be entirely satisfying,
so that a second approach is taken in which the ferrofluid drop is treated as a
\emph{liquid half-sphere} with its own inner flow field. In this case, the
velocity fields and also the sums of viscous and magnetic stresses must be continuous
at the interface between the two liquids. Though the liquid drop cannot be described
as a whole sphere but only as a half-sphere, the resultant drop speed, which no longer
depends on any unknown parameters, represents the experimental data
extremely well. This may indicate that the true flow field in the drop
is mainly restricted to its lower part.
%%%%%%%%%%%%%%%%%%%%%%%%%%%%%%%%%%%%%%%%%%%%%%%%%%%%%%%%%%%%%%%%%%%%%%%%%%%%%%%%%%%%%%
%%%%%%%%%%%%%%%%%%%%%%%%%%%%%%%%%%%%%%%%%%%%%%%%%%%%%%%%%%%%%%%%%%%%%%%%%%%%%%%%%%%%%%

\subsection{Solid sphere}\label{sec_Nav}

The basic hydrodynamic equations are the continuity equation for incompressible fluids

\bq
\nabla\cdot\vv = 0      \label{cont}\,,
\eq

and the stationary Stokes equation

\bq
\mathbf{0} = -\nabla p + \eta\nabla^2\vv        \label{stokes}
\eq

which by eliminating the pressure can be written as

\bq
\nabla^2 \left( \nabla\times\vv \right) = \mathbf{0}\,.         \label{stokesrot}\,
\eq

The velocity field of the non-magnetic liquid bearing the sphere is expanded in
vector spherical harmonics according to Ref.\,\cite{sorokin,MoFe}.
\ref{flowfieldapp} gives the details of this expansion and shows how the
various coefficients occurring in it are determined from the boundary
conditions.

When only one boundary condition is left, namely the requirement that
the dissipating viscous torque compensate for the accelerating
magnetic torque, the velocity components of the flow field below the
sphere still depend on the yet unknown angular velocity $\itom$ with
which the sphere is rotating. The resulting expressions are
(cf. \ref{flowfieldapp}):

\bqa\fl
\frac{v_r}{\itom R}
= \frac{1}{2} \,\frac{\cph\sth}{1+2\frac{\ls}{R}}\left[ 1 - \frac{R^3}{r^3} \right]
                \nonumber\\
+\frac{1}{2}\cph
    \sum_{\stackrel{\ell=3}{\ell\textrm{ \scriptsize odd}}}^\infty \pleins(\cth)
    \frac{R^\ell}{r^\ell}
    \left[ 1 - \frac{R^2}{r^2} \right]
    \frac{(-1)^{\frac{\ell-1}{2}}  (2\ell+1)}{1+(2\ell+1)\frac{\ls}{R}}\cdot
    \frac{(\ell-2)!!}{(\ell+1)!!}
    \label{vr_Nav}
\eqa
and
\bqa\fl
\frac{1}{\itom R}
\left(
\eqalign{\vth \\ \vphi}
\right)
= \frac{1}{2}
    \left(
    \eqalign{ \cph\,\cth \\ -\sph}
    \right)
    \left[ 1 + \frac{R^3}{2r^3} \right]\frac{1}{1+2\frac{\ls}{R}}
    \nonumber\\
+ \frac{1}{2}
    \left(
    \eqalign{ \cph \,\parth \\ -{\sph}/{\sth} }
    \right)     \nonumber\\
    \times
    \sum_{\stackrel{\ell=3}{\ell\textrm{ \scriptsize odd}}}^\infty
    \frac{\pleins(\cth)\, (-1)^{\frac{\ell-1}{2}}}{1+(2\ell+1)\frac{\ls}{R}}\,
    \frac{R^\ell}{r^\ell}
    \left[ (2-\ell) + \ell\frac{R^2}{r^2} \right]
    \frac{(2\ell+1)(\ell-2)!!}{\ell(\ell+1)(\ell+1)!!}     \nonumber\\
+ 2    \left(
    \eqalign{ -\cph/\sth \\ \sph\,\parth}
    \right)
    \sum_{\stackrel{\ell=2}{\ell\textrm{ \scriptsize even}}}^\infty
    \frac{\pleins(\cth)\, (-1)^\frac{\ell}{2}}{1+(\ell+2)\frac{\ls}{R}}\,
    \frac{R^{\ell+1}}{r^{\ell+1}}\,
    \frac{(2\ell+1)(\ell-3)!!}{\ell(\ell+1)(\ell+2)!!}\; .
    \label{vthphi_Nav}
\eqa

The flow field determines the pressure via Stokes' equation (\ref{stokes}).
Straightforward calculation yields

\bqa
\nabla^2\vv &=& \nabla\sum_{\ell,m}\frac{2(2\ell-1)}{(\ell+1)}\frac{\clm}{r^{\ell+1}}\ylm
   = \frac{1}{\eta} \nabla p
\eqa

so that the pressure field is given by

\bqa\fl
p\of = \eta \sum_{\ell,m} \frac{2(2\ell-1)}{(\ell+1)}\frac{\clm}{r^{\ell+1}} \ytp
        \nonumber\\
  = \frac{3}{4}\,\eta\itom \,\frac{\cph \sth}{1+2\frac{\ls}{R}}    \frac{R^2}{r^2}
        \nonumber\\
    + \,\eta\itom \cph \sum_{\stackrel{\ell=3}{\ell\textrm{ \scriptsize odd}}}^\infty
    \frac{\pleins (\cth)\,(-1)^\frac{\ell-1}{2}}{1+(2\ell+1)\frac{\ls}{R}}\,
    \frac{R^{\ell+1}}{r^{\ell+1}}\,
    \frac{4\ell^2-1}{\ell+1}\,\frac{(\ell-2)!!}{(\ell+1)!!} \,.
\eqa

\subsection{Viscous torque}

The viscous torque acting on the lower half-sphere is
gained from the tangential viscous forces

\bq
\de\bi{F}_\mathrm{tang} = [\sigrt\etheta + \sigrp\ephi] R^2\dphi\,\dth\sth
\eq

according to

\bq
\de\bi{T}_\mathrm{vis} = \bi{R}\times\de\bi{F}_\mathrm{tang}(r=R)
\eq

with the tangential components of the viscous stress tensor $\sigrt$ and $\sigrp$ as
defined in \cite{LLhydro}. Integration over the lower half-sphere
$0\leq\vartheta\leq\pi/2$, $0\leq\varphi\leq 2\pi$ yields the dimensionless viscous
torque in $y$-direction
\bqa\fl
\frac{-T_\mathrm{vis}}{\pi\eta\,\itom R^3} =
   \frac{3}{2}\, \frac{1}{1+2\frac{\ls}{R}}
 + \lim_{N\rightarrow\infty}
   \sum_{\stackrel{\ell=3}{\ell\textrm{ \scriptsize odd}}}^N
   \frac{(2\ell+1)^2}{1+(2\ell+1)\frac{\ls}{R}} \left[ \frac{(\ell-2)!!}{(\ell+1)!!}
   \right]^2  \nonumber\\
 + \lim_{N\rightarrow\infty}
   \sum_{\stackrel{\ell=2}{\ell\textrm{ \scriptsize even}}}^N
   \frac{4(2\ell+1)(\ell+2)}{1+(\ell+2)\frac{\ls}{R}}
   \left[ \frac{(\ell-3)!!}{(\ell+2)!!} \right]^2 \,.
   \label{viscous_torque_Nav}
\eqa

When the doublefactorials in (\ref{viscous_torque_Nav}) are transformed to single
factorials and Stirling's approximation

\bq
\ell!\approx \sqrt{2\pi\ell}\,\ell^\ell\ehoch{-\ell},\qquad \ell\gg 1
\eq

is employed, it can be shown that the terms for large $\ell$ in the infinite series
give in leading order
\bqa
\frac{2}{\pi \ell^2}\frac{R}{\ls}, &\qquad \textrm{for } \ls>0 \\
\frac{4}{\pi \ell} , &\qquad \textrm{for } \ls=0.
\eqa

While $\sum_{\ell=1}^\infty \ell^{-2}$ is a convergent series, $\sum_{\ell=1}^\infty
\ell^{-1}$ diverges logarithmically, so here the necessity of the slip condition
becomes manifest.

Looking at the solution (\ref{vr_Nav}), (\ref{vthphi_Nav}) for the velocity field, it
becomes clear that the field is only changed significantly near the contact line or,
more generally, near the sphere surface: since $\ls \ll R$, the terms with small
$\ell$ hardly deviate from those for no-slip. Only when $\ell{\ls}/{R}$ exceeds the
order unity, the factors containing $\ls$ become important. Each term is made smaller,
and the more so the greater $\ell$ becomes and, of course, the greater the slip length.
On the other hand, the terms with large $\ell$, \emph{i.e.}, those
which are influenced by the slip condition, are negligible when $r\gg R$. So the results
with and without slippage would not be distinguishable far enough from the contact line.

Figures \ref{vth_Nav} and \ref{sigrth_Nav} illustrate the influence of slippage in the
relevant region near $\vartheta=\pi/2$ for expansion orders 99 and 100, respectively.
Where there is a steep descend in the dependence of $\vth^{(100)}(r=R)$ on $\vartheta$
and therefore a large corresponding tangential viscous stress $\sigrt^{(99)}(r=R)$
for $\ls=0$, the curves are considerably smoothed out when the fluid is allowed to slip.

\begin{figure}[tb]
\begin{minipage}{0.4\textwidth}
  \caption{Influence of slipping on $\vth^{(N)}(R)$ over $\vartheta$ for $N=100$.
                Both the oscillations and the steep descent to zero are considerably
                smoothed out when a finite slip length is taken into account.\\[1cm]}
  \label{vth_Nav}
  \caption{Influence of slipping on the relevant stress component $\sigrt^{(N)}(R)$ over
                $\vartheta$ for $N=99$. The greater the slip length, the more are the
                oscillations damped, \emph{i.e.}, the more is the stress relieved.}
  \label{sigrth_Nav}
\end{minipage}
\begin{minipage}{0.5\textwidth}
  \includegraphics[width=7cm]{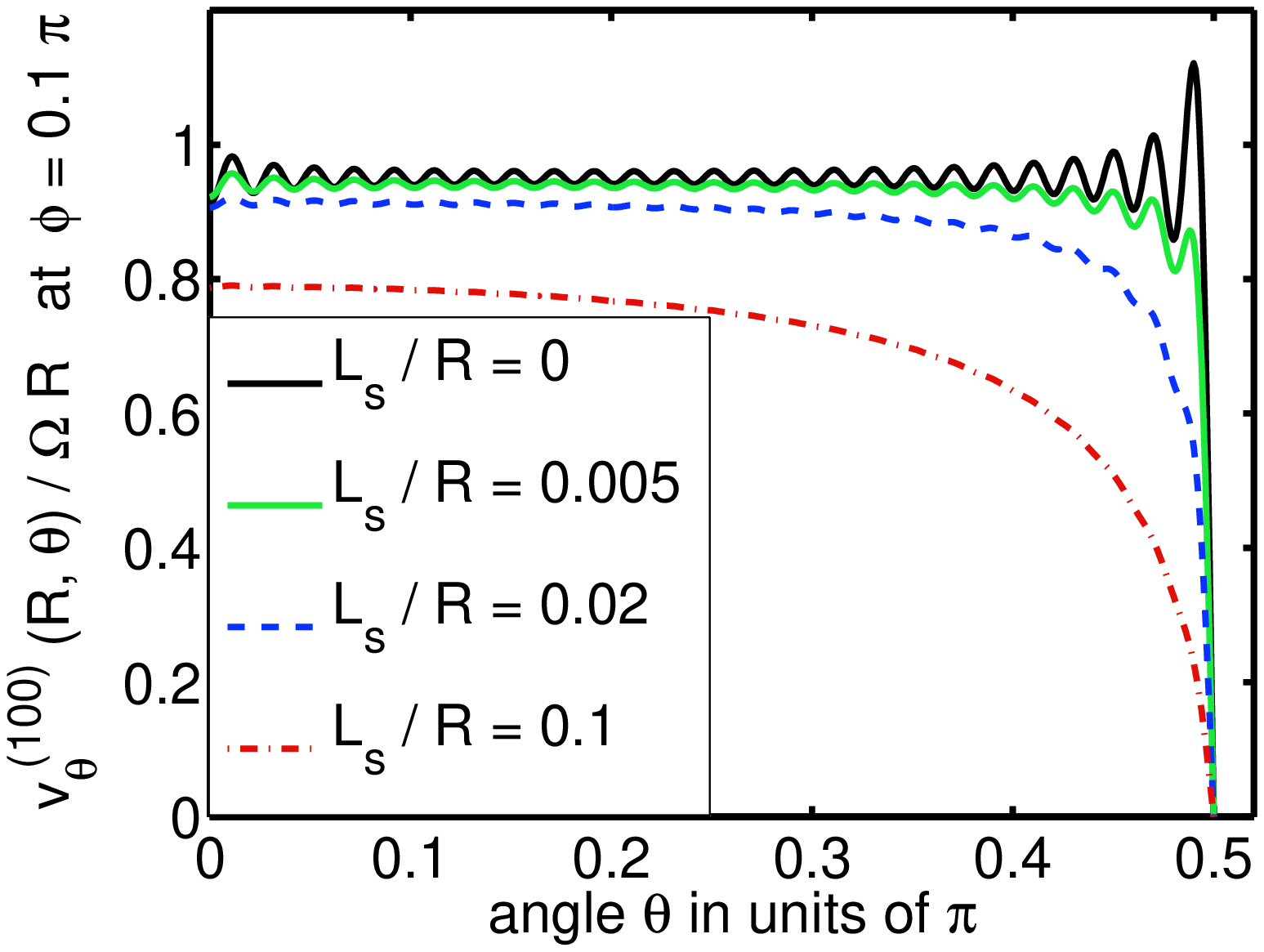}\\[0.5cm]
  \includegraphics[width=7cm]{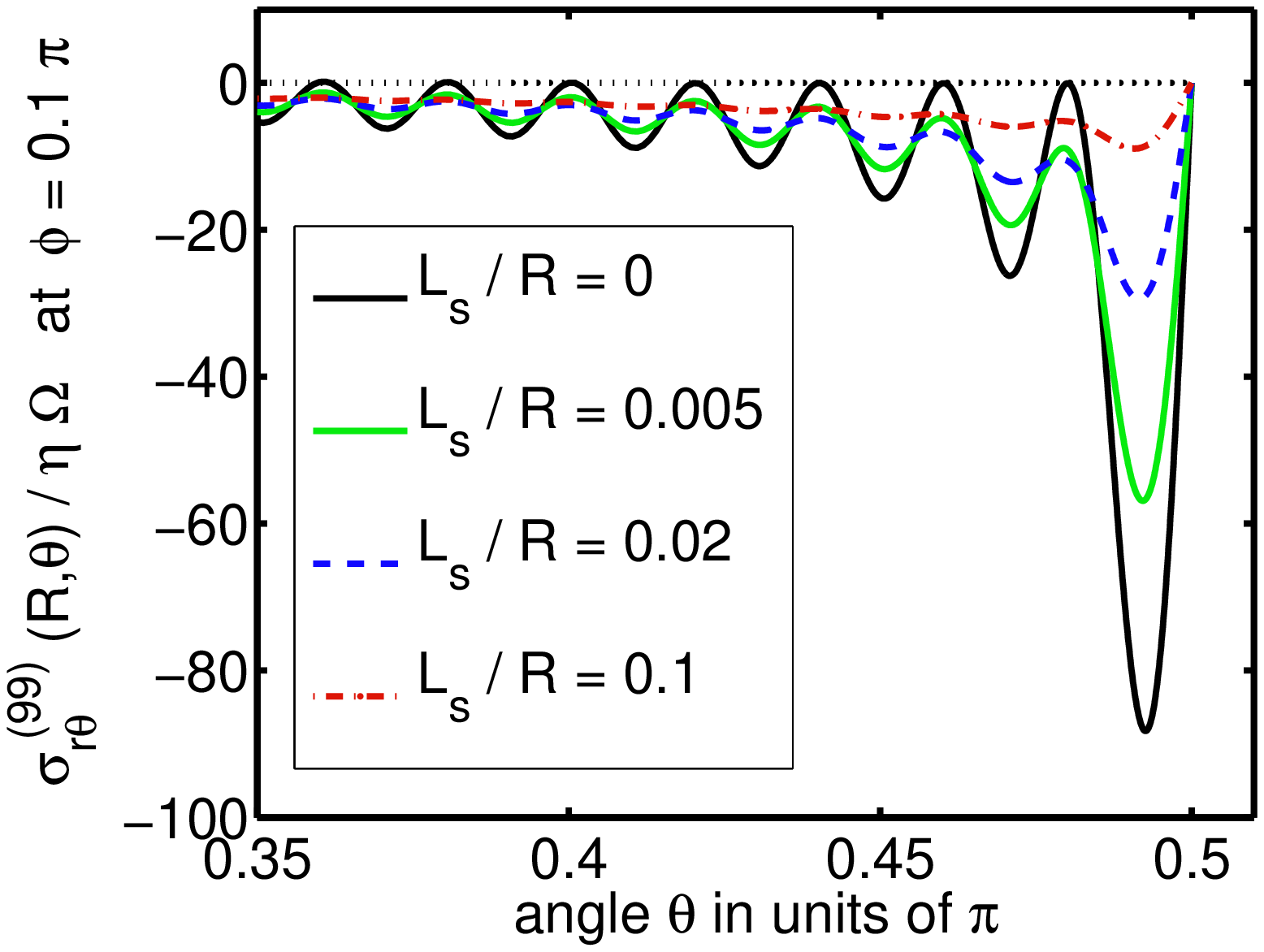}\fl
  \end{minipage}\hspace{1.5cm}
\end{figure}

%%%%%%%%%%%%%%%%%%%%%%%%%%%%%%%%%%%%%%%%%%%%%%%%%%%%%%%%%%%%%%%%%%%%%%%%%%%%%%%%%%%%%%%

\subsection{Magnetic torque}

In order to obtain an expression for the angular velocity $\itom$, we utilize
the fact that the viscous torque (\ref{viscous_torque_Nav}) must compensate for
the magnetic torque which is calculated now.

The vector of the applied magnetic field rotates within the $xz$-plane,
generating a torque in $y$-direction, so that the external magnetic field is denoted by

\bq
\bi{G} = \real\{\mathbf{\hat{\bi{G}}}\}
, \qquad \mathbf{\hat{\bi{G}}} = G \,\ehoch{\ii\omega t}(-\ii\eix+\ez)
\eq

with $\omega=2\pi f$ being the rotation frequency of the field and

\bq \chi = \chi\prime - \ii\chi\prime\prime = \chi(f) \eq

the frequency dependent magnetic susceptibility of the sphere. Concerning the amplitude
of the magnetic field, the susceptibility is assumed to be a constant.

The sphere is supposed to be magnetized homogeneously, having the overall magnetization
(see for example \S\S\ 8 and 29 in \cite{LLmag})

\bq
\bi{M} = \real\{\mathbf{\hat{\bi{M}}}\}
, \qquad \mathbf{\hat{\bi{M}}} = \frac{G\chi}{1+\frac{\chi}{3}}
\,\ehoch{\ii\omega t}(-\ii\eix+\ez)
\eq

so that the magnetic torque acting on it in the stationary state is given by \cite{rosensweig1985}

\bq
\bi{T}_\mathrm{mag} = \mu_0 V\ey\,(M_z G_x-M_x G_z)
= \frac{4\pi}{3}R^3 \frac{ \mu_0G^2\chi\prime\prime}%
{(1+\frac{\chi\prime}{3})^2+(\frac{\chi\prime\prime}{3})^2}\,\ey \,.
\eq

This must compensate for the viscous torque

\bq
\bi{T}_\mathrm{vis} = -\pi\eta\,\itom R^3 \, \Sigma(\ls)\,\ey\,.
\label{Om_from_torque}
\eq

Here, the right-hand side of (\ref{viscous_torque_Nav}) is abbreviated
by $\Sigma(\ls)$, reminding us that it includes an infinite series
which depends on the slip length and cannot be computed analytically
in closed form. The equality of viscous and magnetic torques poses the
last boundary condition which makes sure that the rotational and,
consequently, also the translational motion of the sphere be not
accelerated, and finally gives the rotation frequency of the sphere:

\bqa
\itom = \frac{4}{3}
        \frac{\mu_0 G^2 \chi\prime\prime}%
        {\left[(1+\frac{\chi\prime}{3})^2+(\frac{\chi\prime\prime}{3})^2\right]
        \eta\,\Sigma(\ls)}
\equiv \frac{8}{3}\frac{\mathfrak{M}}{\eta\,\Sigma(\ls)}
\eqa

The speed with which the sphere advances on the fluid surface is given by the negative
of the velocity field at $r\to\infty$. In this limit, only the $(\ell=1)$-terms remain
and the corresponding factor from the Navier slip condition
can be neglected because of $\ls\ll R$:

\bqa
\vv_{\mathrm{drop}}
= -\frac{\itom R}{2}\left(
\eqalign{\sth\cph\\ \cth\cph\\ -\sph}
\right)         = -\frac{4}{3}\frac{\mathfrak{M}R}{\eta\,\Sigma(\ls)}\,\eix
\label{solid_sphere_speed}
\eqa
%%%%%%%%%%%%%%%%%%%%%%%%%%%%%%%%%%%%%%%%%%%%%%%%%%%%%%%%%%%%%%%%%%%%%%%%%%%%%%%%%%%%%%%
%%%%%%%%%%%%%%%%%%%%%%%%%%%%%%%%%%%%%%%%%%%%%%%%%%%%%%%%%%%%%%%%%%%%%%%%%%%%%%%%%%%%%%%

\subsection{Fluid (half-)sphere}\label{sec_fluid}

Although a definite result has been obtained for the speed of the magnetic sphere,
it cannot be compared to experimental data so easily.
It still depends on an unknown parameter, the slip length $\ls$, which cannot simply
be treated as a fit parameter. Due to the very weak dependence of the viscous torque
on the expansion order, it poses a formidable numerical problem to obtain the slip
length for a given torque, so it would be of advantage to obtain an expression for the
drop speed that does not depend on such a parameter.

In addition, one could expect a model containing a \textit{liquid drop} to be
more realistic than one with a {\it solid sphere}. For these reasons the ferrofluid drop
is now considered liquid, though still spherical, being also subject to the hydrodynamic
equations like the surrounding liquid. The Navier slip condition is replaced by the
condition of continuous velocities and stresses at the interface between the two liquids.
All other boundary conditions remain as before, including the addition of the mirror
image. As a consequence of the requirement of a flat "surface" ($\vth=0$ at $z=0$ for
all $r>R$), it is not possible to obtain a spherical inner (i) velocity field:
$\vth^{\mathrm{(i)}}$ is rendered zero within the whole section $z=0$ when the
corresponding outer (o) component $\vth^{\mathrm{(o)}}$ is demanded to vanish on the
whole contact circle $z=0$, $r=R$.

However, when the boundary conditions are posed in analogy to the previous
section, a flow field is obtained which proves to be very useful. As the field
becomes completely horizontal within the plane of symmetry, it is suggested
that only the lower half-sphere is identified with the ferrofluid drop,
\emph{i.e.}, after solving the mirror image set-up, the whole upper half-space
is neglected, resulting in the flow field displayed in figure \ref{velc_1}.

\begin{figure}[tb]
  \centering\includegraphics[width=11cm]{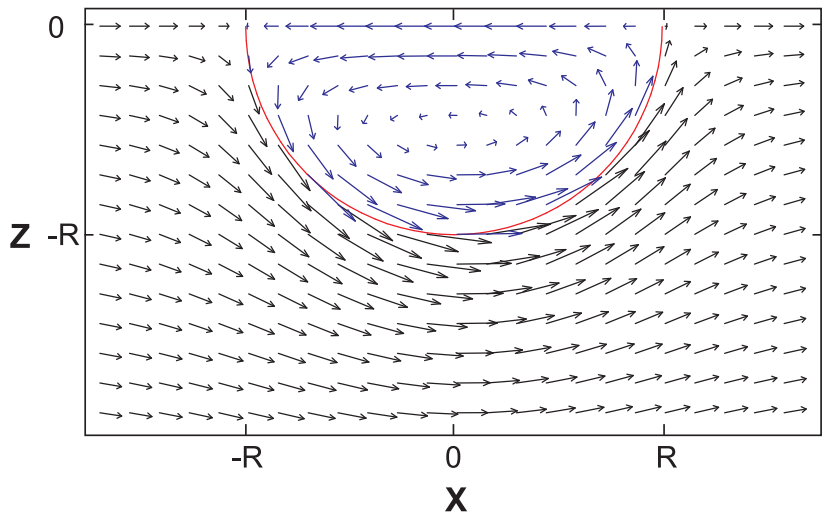}
  \caption{Flow field of the liquid half-sphere within the plane \mbox{$y=0$}.\\}
  \label{velc_1}
\end{figure}

The same differential equations (\ref{cont}), (\ref{stokesrot}) and ansatz
(\ref{expan_vr}), (\ref{expan_real}) together with the requirement that the velocity be
finite at $r=0$ yield for the radial functions of the inner velocity field ($\ell>0$):

\bqa
f_{00}^{\mathrm{(i)}}(r) \equiv 0   \\
\iflm(r) = \qlm r^{\ell+1} + \Blm r^{\ell-1}    \\
\iglm(r) = \frac{\ell+3}{\ell(\ell+1)}\,\qlm r^{\ell+1} + \frac{\Blm}{\ell}\, r^{\ell-1}
\\
\ihlm(r) = \pplm r^\ell
\eqa

Starting point for the velocity components of the surrounding liquid are again the
radial functions (\ref{hlm_a}) - (\ref{glm_cd}).
For simplicity it is still assumed that the drop remains spherical, \emph{i.e.},

\bq
v_r^{(\mathrm{i})}(R) = v_r^{(\mathrm{o})}(R) = 0 \qquad \forall\,\vartheta,\varphi,
\eq

instead of demanding that the normal stresses be continuous at $r=R$.

As mentioned above, the tangential components $\vth$ and $\vphi$
must be continuous. Due to the orthogonalities (\ref{orth2}) and (\ref{orth3}) this
condition reduces to the radial functions $\glm$ and $\hlm$ being continuous.

Furthermore, the tangential forces must cancel out at every point on the
spherical interface so that the tangential stresses are pointwise continuous.
The latter consist of viscous stresses
$\sigma_{r\vartheta/\varphi}^{\mathrm{(vis)}}\equiv\sigma_{r\vartheta/\varphi}$
and magnetic stresses \cite{shliomis}

\bqa
\sigma_{ij}^{(\mathrm{mag})} = \mu_0 H_i H_j - \frac{\mu_0}{2}\,H_i H_j \delta_{ij}
 + \frac{\mu_0}{2}\,(M_i H_j - M_j H_i)   \label{mag_stress_tensor}
\eqa

where $i,j=x,y,z$ and the local magnetic field is given by

\bq
\bi{H}=\real\{\mathbf{\hat{\bi{H}}}\}, \qquad
\mathbf{\hat{\bi{H}}} = \frac{G}{1+\frac{\chi}{3}}\, \ehoch{\ii\omega t}\,(-\ii\eix+\ez),
\eq

assuming a linear magnetization law

\bq
\mathbf{\hat{\bi{M}}} = \chi\mathbf{\hat{\bi{H}}}.
\eq

The quantities $\bi{M}$, $\bi{G}$, $\chi$, and $\mathfrak{M}$ are defined as in the
previous section. For the condition of continuous tangential stresses, the symmetric
part of the magnetic stress tensor (\ref{mag_stress_tensor}) need not be considered
since it is the same on both sides of the interface due to the usual boundary
conditions for $\bi{H}$.

The antisymmetric part, on the other hand, is the crucial one which leads to the
propagation of the drop. It shall be denoted by $\sigma_{ij}^{(\mathrm{m})}$. Because
of antisymmetry in addition to $\hat{{H}}_y=\hat{M}_y=0$, only one independent
cartesian component is left:

\bq
{\sigma_{xz}^{(\mathrm{m})}} =
    -\frac{\mu_0}{2}\,\frac{G^2 \chi\prime\prime}%
    {\left(1+\frac{\chi\prime}{3}\right)^2+\left(\frac{\chi\prime\prime}{3}\right)^2}
= -\mathfrak{M}
\eq

This gives the tangential magnetic stresses
\bqa
{\sigrt^{(\mathrm{m})}} = \mathfrak{M}\cph\\
{\sigrp^{(\mathrm{m})}} = -\mathfrak{M}\cth\sph.
\eqa

Now the boundary condition reads

\bq
F_{\vartheta/\varphi}^{\mathrm{(m)}}(R)
    = F_{\vartheta/\varphi}^{\mathrm{(vis,i)}}(R)
    + F_{\vartheta/\varphi}^{\mathrm{(vis,o)}}(R)       \label{forces_compensate}
\eq

because the accelerating magnetic force must be compensated by the viscous ones.
With $F_{\vartheta/\varphi}=\sigma_{r\vartheta/\varphi}\,\mathbf{n}\!\cdot\!\er$ and
the convention that the surface normal $\mathbf{n}=+\er$ for a force that acts on the
\textit{outer} surface and $\mathbf{n}=-\er$ for a force that acts on the \emph{inner}
surface, this yields in terms of stresses

\bq
\sigma_{r\vartheta/\varphi}^{\mathrm{(m)}}
    + \sigma_{r\vartheta/\varphi}^{\mathrm{(vis,o)}}(R)
    - \sigma_{r\vartheta/\varphi}^{\mathrm{(vis,i)}}(R) = 0     \label{sum_of_sigmas}
\eq

for all $\vartheta,\varphi$.
As before, the viscous force in $x$-direction must vanish.
The resulting expressions of the components of inner and outer flow field
are given explicitely in \ref{thirdapp}.

Again, the speed of the drop in the laboratory frame is obtained by evaluating the
negative of the outer velocity field at $r\rightarrow \infty$, giving

\bq
\vv_\mathrm{drop}^\mathrm{liq} =
  - \frac{1}{2} \frac{\mathfrak{M}R}{2\eta^{(\mathrm{o})}+3\eta^{(\mathrm{i})}}\,\eix\,.
  \label{drop_speed}
\eq

Although this result looks very similar to the one obtained in the previous section,

\bq
\vv_\mathrm{drop}^\mathrm{sol} = - \frac{4}{3}\frac{\mathfrak{M}R}{\eta^{\mathrm{(o)}}
\Sigma(\ls)}\,\eix\,, \label{drop_speed_Nav}
\eq

it clearly has two advantages. First, it purely consists of parameters that are
experimentally measurable or tunable (sphere radius $R$, viscosities $\eta$, and via
$\mathfrak{M}$ susceptibility $\chi$ and external magnetic field amplitude $G$).
Second, there is no need of calculating
numerically an infinite sum.

Since no singularity has occurred in the scope of the calculations for the
liquid sphere, it can be compared to a model where slipping is taken into
account. The stresses which diverge within the framework of the very rigid
no-slip condition are relieved both when the surrounding fluid is allowed to
slip over the solid and when the solid is replaced by an elastic or, as in our
case, viscous medium. Indeed, the crucial viscous stress component
$\sigrt^{\mathrm{liq}}\equiv\sigrt^{\mathrm{(vis,o)}}$ is essentially identical
to the one obtained from the velocity field with Navier slip, the only
differences being constant factors, at least when $\ell\gg1$:

\bqa\fl
\frac{\sigrt^{\mathrm{sol}}(R)}{\mathfrak{M}/\Sigma(\ls)} =
        -\frac{4}{1+2{\ls}/{R}} \,\cph\cth \nonumber\\
- \frac{8}{3} \cph \sum_{\stackrel{\ell=3}{\ell\textrm{ \scriptsize odd}}}^\infty
              \parth \pleins(\cth)\,
              \frac{(-1)^\frac{\ell-1}{2}}{{1}/({2\ell+1})+{\ls}/{R}}\cdot
\frac{(2\ell+1)(\ell-2)!!}{\ell(\ell+1)(\ell+1)!!}
              \nonumber\\
+ \frac{16}{3} \frac{\cph}{\sth}
              \sum_{\stackrel{\ell=2}{\ell\textrm{ \scriptsize even}}}^\infty
              \pleins(\cth)\,
              \frac{(-1)^{\frac{\ell}{2}}}{{1}/{(\ell+2)}+{\ls}/{R}}\cdot
\frac{(2\ell+1)(\ell-3)!!}{\ell(\ell+1)(\ell+2)!!}
\eqa

\bqa\fl
\frac{\sigrt^{\mathrm{liq}}(R)}{\mathfrak{M}} =
        -\frac{3}{2}\,\frac{1}{2 + 3{ \eta^{\mathrm{(i)}}}/{\eta^{\mathrm{(o)}} }}
        \,\cph\cth  \nonumber\\
- \frac{2\cph}{1+{ \eta^{\mathrm{(i)}}}/{\eta^{\mathrm{(o)}} }}
        \sum_{\stackrel{\ell=3}{\ell\textrm{ \scriptsize odd}}}^\infty
        \parth \pleins(\cth)\, (-1)^\frac{\ell-1}{2} \,
\frac{(2\ell+1)(\ell-2)!!}{\ell(\ell+1)(\ell+1)!!}      \nonumber\\
+2\,\frac{\cph}{\sth}
        \sum_{\stackrel{\ell=2}{\ell\textrm{ \scriptsize even}}}^\infty
        \frac{\pleins(\cth)\,(-1)^\frac{\ell}{2}}%
          { 1 + ({\ell-1})/({\ell+2})\cdot{\eta^{\mathrm{(i)}}}/{\eta^{\mathrm{(o)}}} }
\cdot\frac{(2\ell+1)(\ell-3)!!}{\ell(\ell+1)(\ell+2)!!}
\eqa
%%%%%%%%%%%%%%%%%%%%%%%%%%%%%%%%%%%%%%%%%%%%%%%%%%%%%%%%%%%%%%%%%%%%%%%%%%%%%%%%%%%%%%%
%%%%%%%%%%%%%%%%%%%%%%%%%%%%%%%%%%%%%%%%%%%%%%%%%%%%%%%%%%%%%%%%%%%%%%%%%%%%%%%%%%%%%%%

\section{Comparison of experimental and theoretical results}\label{sec_comp}

The main result of the theory for the drop speed (\ref{drop_speed}) reads
explicitly

\bq
v_\mathrm{drop}^\mathrm{liq} =
  - \frac{R}{4}\,\frac{\mu_0 G^2 }{2\eta^{(\mathrm{o})}+3\eta^{(\mathrm{i})}}
  \cdot\frac{\chi\prime\prime}{(1+\frac{\chi\prime}{3})^2+(\frac{\chi\prime\prime}{3})^2}
  .  \label{drop_speed_expl}
\eq

\begin{figure}[tb]
\begin{minipage}{0.4\textwidth}
  \caption{The drop speed in dependence of the magnetic field amplitude $G$
  for {$f=0.8$\,kHz} and {$V=5\,\tcmu$l}. The blue circles mark the
  measured data, the red line gives the theoretical curve according to
  (\ref{drop_speed_expl}), taking into account the proper material values.}
  \label{v_over_field}
\end{minipage}\hspace{0.8cm}
\begin{minipage}{0.6\textwidth}
  \includegraphics[width=8cm]{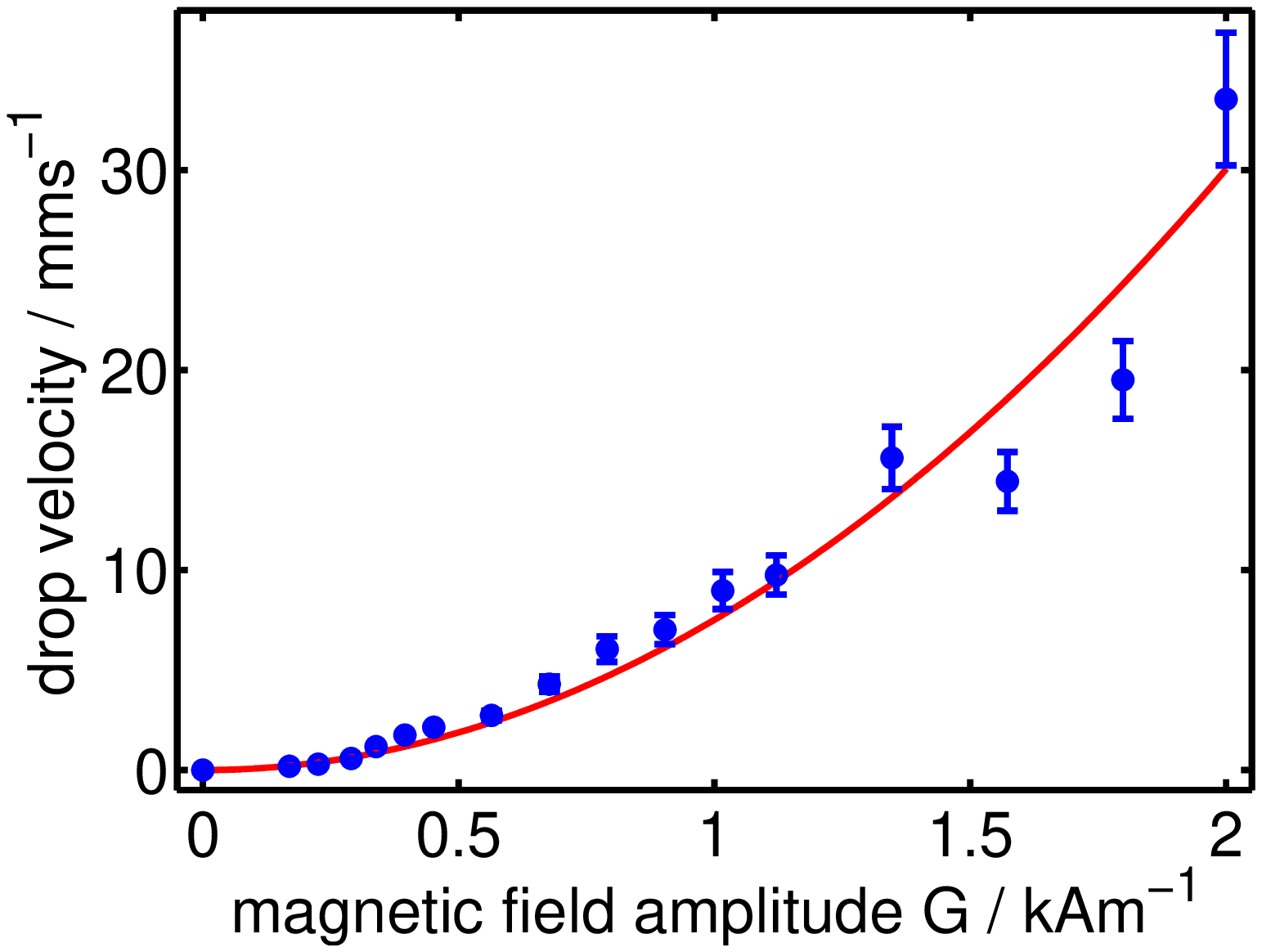}
\end{minipage}
\end{figure}
\begin{figure}[htb]
\begin{minipage}{0.4\textwidth}
\caption{Drop velocity versus drop radius for an alternating magnetic far field
with $G = 0.844$\,kA/m and $f=0.8$\,kHz. The blue dots mark the experimental
results, the solid line the theoretical outcome.} \label{v_over_vol}
\end{minipage}\hspace{0.8cm}
\begin{minipage}{0.6\textwidth}
  \includegraphics[width=8cm]{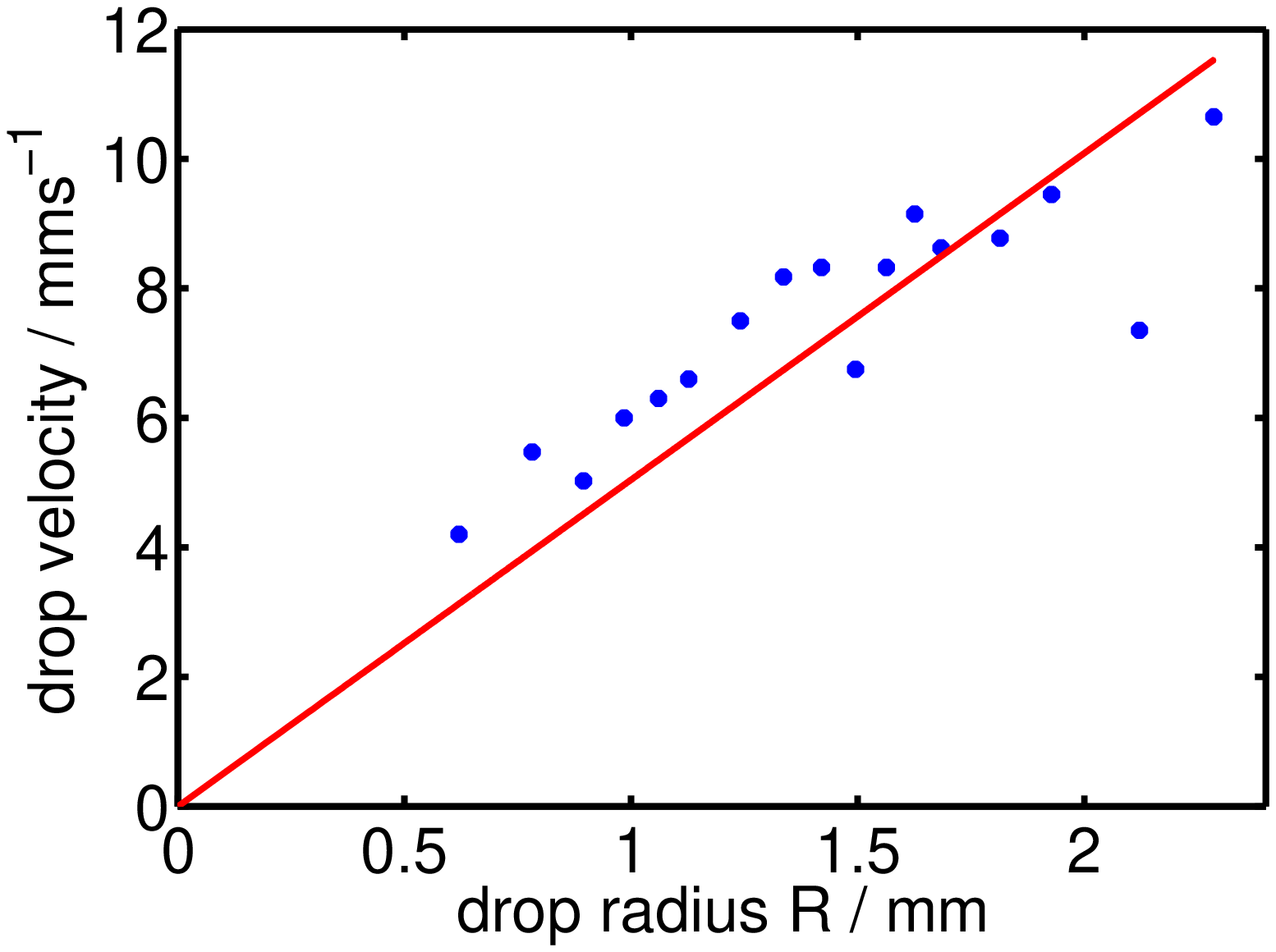}
\end{minipage}
\end{figure}

It can be well compared with the data obtained in the experiments. They have
been measured following the procedure described in section~\ref{sec_experiment}.
Figure~\ref{v_over_field} presents a plot of the drop velocity versus the
magnetic field amplitude $G$ for a driving frequency of $f=0.8$\,kHz. We have
put a droplet of volume $V=5\,\tcmu$l, corresponding to a sphere of radius
$R\approx 1.1$\,mm, on top of the liquid layer. The measured velocities (marked
by full circles) show a monotonous increase with $G$. The solid line gives the
values of (\ref{drop_speed_expl}), taking into account the viscosities of
the ferrofluid,  $\eta^\mathrm{(i)}=5.4$\,mPa\,s and of the liquid below, which
amounts to $\eta^\mathrm{(o)}=1.27$\,mPa\,s. The driving frequency enters into
expression (\ref{drop_speed_expl}) only via the real and imaginary part of the
magnetic susceptibility which were determined as $\chi\prime=4.66$ and
$\chi\prime\prime=3.25$, respectively, for the given frequency
(cf.~figure~\ref{fig:chi}). As can be seen,
the values for the liquid half-drop solution represent the given experimental
data extremely well.

In a series of measurements different drops with a volume ranging from $1$ to
$50\,\tcmu$l were investigated. For comparison with theory we assume a
spherical symmetry and estimate the drop radius $R$ from the drop volume $V$.
As shown in figure~\ref{v_over_vol}, the measured drop velocity increases with
the radius of the drops. The solid line marks the result of (\ref{drop_speed_expl})
for an amplitude of $G=0.884$\,kA/m, as set in the experiment.
Again we find a quantitative agreement of the half-drop solution with the
experimental data.

\begin{figure}[htb]
\centering
\includegraphics[width=0.6\textwidth]{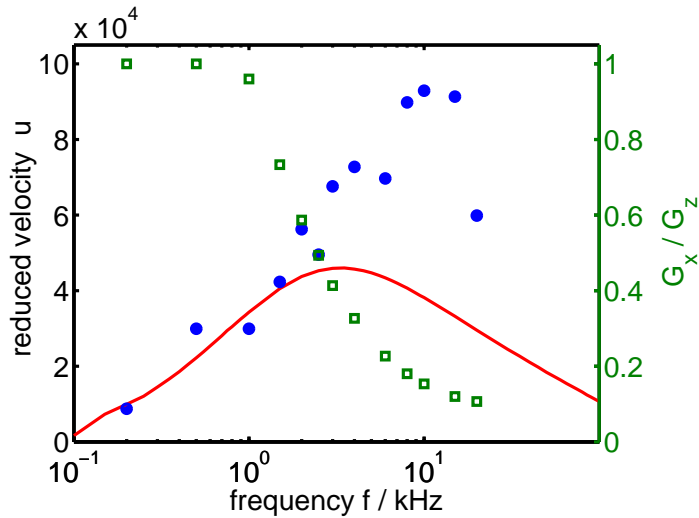}
  \caption{\ Frequency dependence of the reduced drop velocity $u$ for $V=5\,\tcmu$l.
  The full circles mark the experimental data, the solid line gives
  the theoretical curve where the measured frequency dependence of
  $\chi\prime(f)$ and $\chi\prime\prime(f)$ has been plugged in. For all data
  $G_z$ was fixed to 0.844\,kA/m, but $G_x$ was decreasing with increasing $f$.
  The green open squares are indicating the actual ratio $G_x/G_z$.}
  \label{v_over_freq}
\end{figure}

As a further parameter the driving frequency $f$ was varied in the experiment.
When the frequency dependence of the drop speed was determined, the vertical
field was fixed at $G_z=0.844$\,kA/m. However, the frequency dependent
inductance of the outer coils did not permit to keep $G_x$ at this value for
the whole range of frequencies (the ratio $G_x/G_z$ is indicated at the
r.h.s.~of figure~\ref{v_over_freq}). In order to obtain a magnitude which is
independent of $G$, we introduce the reduced velocity
\begin{equation}
u = v_\mathrm{drop}^\mathrm{liq} \frac{\eta^{\mathrm{(i)}}}{R  \mu_0  G_x G_z},
\label{eq:pump.u}
\end{equation}
where  $G_x$ denotes the horizontal and $ G_z$ the vertical field amplitude.
Within the linear regime this quantity should be independent of the choice of
the amplitudes. Figure~\ref{v_over_freq} shows an increase of the reduced drop
velocity (marked by solid circles) up to a maximum at $f=10$\,kHz. The
theoretical values (solid line) stem from (\ref{drop_speed_expl}), where the
material parameters and the measured frequency dependence of the complex
susceptibility $\chi\prime(f) + i \chi\prime\prime(f)$, as presented in
figure~\ref{fig:chi}, have been utilized. In order to be able to compare the
predictions with the experimental results, $v_\mathrm{drop}^\mathrm{liq}$ is
scaled according to (\ref{eq:pump.u}). We observe a good agreement up to a
frequency of about $f=1.5$\,kHz. Beyond that point, the theoretical curve
deviates from the experimental results. The former shows a maximum at about
$f=3.5$\,kHz, while the measured velocity is largest at $f=10$\,kHz, and the
maximum values differ by a factor of two.

%%%%%%%%%%%%%%%%%%%%%%%%%%%%%%%%%%%%%%%%%%%%%%%%%%%%%%%%%%%%%%%%%%%%%%%%%%%%%%%%%%%%%%%%%%%%%%%%%%%%%%%%%%%%%%%%%%%%%%%%%%%%%%%%%%%%%%%%%%%
%%%%%%%%%%%%%%%%%%%%%%%%%%%%%%%%%%%%%%%%%%%%%%%%%%%%%%%%%%%%%%%%%%%%%%%%%%%%%%%%%%%%%%%%%%%%%%%%%%%%%%%%%%%%%%%%%%%%%%%%%%%%%%%%%%%%%%%%%%%

\section{Discussion and Conclusion}\label{sec_conc}

The measured propagation velocity of the droplet shows a parabolic dependence
from the magnetic field amplitude, and a linear dependence from the radius of
the droplet. Both experimental observations are quantitatively described by the
liquid half-drop solution, without any free fitting parameter. The theory just
needs the magnetic field amplitude, the complex susceptibility and the
viscosities of both fluids (\emph{i.e.}, the ferrofluid and the liquid layer).
Taking into account the over-simplifying assumption of a half-spherical drop,
the theory describes the experimental data remarkably well for driving
frequencies up to $1.5$\,kHz.

For higher driving frequencies, however, (cf.~figure~\ref{v_over_freq}) a
discrepancy between experiment and theory of up to 100\,\% is observed. This
discrepancy may have several origins. Firstly, due to experimental
characteristics, the rotating magnetic field becomes elliptical. Following
\cite{lebedev2003}, the nonlinear effects of an elliptical field are
expected to diminish the flow within the droplet. This, however, does not
explain our experimental data, which overcome the predictions by theory. Of
course our experimental situation differs from that of \cite{lebedev2003}
where an elliptical drop can freely rotate in the horizontal plane.  In our
case the horizontal surface is pinning a free rotation of an elliptical droplet
in the vertical plane.

Secondly, for higher driving frequencies the liquid-liquid interface of a fully
immersed drop develops spikes and resembles a "spiny starfish", as reported in
Refs.\,\cite{bacri1994,lebedev2003}. This may also happen for the lower part of
our half-immersed, swimming drop. A complex interface of the two liquids may
enhance the interaction in between the fluids and thus increase the propulsion
-- similar to a paddle wheel of a Mississippi steam boat. This can of course
not be covered by the simplifying model ansatz. The shape and dynamics of the
liquid-liquid interface shall be studied in forthcoming experiments.

The main achievement of the article is that rotating fields can transport
ferrofluidic drops. Our experimental results can be quantitatively explained
without any free fitting parameters.

Moreover the theory gives an explicit solution of the flow fields both for a
rotating solid magnetic sphere and a spherical ferrofluid drop which both are
half-way immersed in a liquid. The similarity of the final results of both
cases demonstrates the equivalence of Navier slip at a solid surface on the one
hand and the continuity of tangential stresses at a fluid-fluid boundary on the
other hand.

For a quantitative description of "magnetic pumping" by means of a rotating
field  a droplet is more suitable than a plain ferrofluidic layer
\cite{krauss2006}. For the droplet one does not need any tracer particles (the
droplet is its own tracer), and the demagnetization factor of an elliptical
droplet is well defined.

Future experiments shall unveil whether the half-drop model works also in the
pico-liter range. Here the dimensioning of droplets is very precise (see e.g.
Ref.\,\cite{thorsen2001}) and their position may be detected by magnetic
sensors \cite{pekas2004}. Taking advantage of (\ref{drop_speed_expl}) one may
even select the size of the generated droplets by their speed.

We propose that the controlled transport of small amounts of liquid to any
desired position on top of a liquid two dimensional layer is a promising
technique for microfluidic applications. There ferrofluidic drops are commonly
manipulated utilizing local field gradients, which are locally created by
embedded wires \cite{pamme2006} or planar coils \cite{nguyen2006}. In contrast,
our driving technique yields a constant drop velocity globally, i.e.~on the
complete surface.

\begin{ack}
The authors would like to thank Jens Eggers and Thomas Fischer for valuable discussions concerning the theoretical modelling. In addition they thank Norbert Buske for drawing their
attention to the per-fluorinated liquid, Nina Matoussevitch for her excellent
magnetic fluid, and Marit {\O}verland for experimental support. Moreover
R.K.~and R.R.~gratefully acknowledge financial support from the collaborative
research center SFB 481 via project B9.
\end{ack}
%%%%%%%%%%%%%%%%%%%%%%%%%%%%%%%%%%%%%%%%%%%%%%%%%%%%%%%%%%%%%%%%%%%%%%%%%%%%%%%%%%%%%%%%%%%%%%%%%%%%%%%%%%%%%%%%%%%%%%%%%%%%%%%%%%%%%%%%%%%
%%%%%%%%%%%%%%%%%%%%%%%%%%%%%%%%%%%%%%%%%%%%%%%%%%%%%%%%%%%%%%%%%%%%%%%%%%%%%%%%%%%%%%%%%%%%%%%%%%%%%%%%%%%%%%%%%%%%%%%%%%%%%%%%%%%%%%%%%%%

\begin{appendix}

\section{Explicit computation of the flow field below the solid sphere}
\label{flowfieldapp}

The velocity field is expanded in vector spherical harmonics according to
\cite{sorokin,MoFe}

\bqa\fl
\vv\of
= \mysum\Bigg\{ \er\flm(r) + \glm(r)r\nabla + \hlm(r)\vecr\times\nabla \Bigg\}\ytp
 \label{expan1}
\eqa

with the normalized spherical harmonics $\ylm$ and the Legendre functions $\plm$
as defined in \cite{cohentann} for $\ell \geq 0$ and $0\leq m\leq \ell$:

\bqa
\ytp &= (-1)^m\sqrt{\frac{2\ell +1}{4\pi}\frac{(\ell-m)!}{(\ell+m)!}}
 \times\,\ehoch{\ii m\varphi}\pth \\
&\equiv \klm\,\ehoch{\ii m\varphi}\pth          \label{klmdef}
\eqa

\bq
\ytpmi = (-1)^m\ytpst   \label{ylm_minus}
\eq

\bq
\pth = \frac{(-1)^\ell}{2^\ell\ell!}(\sth)^m  \dlm(\sth)^{2\ell}    \label{plm_mnotneg}
\eq

When the expansion (\ref{expan1}) is put into (\ref{cont}) and (\ref{stokesrot}),
these \emph{partial} differential equations for the \emph{vector} $\vv$
are transformed to \emph{ordinary} differential equations for the
\emph{scalar} radial functions $\flm$, $\glm$, and $\hlm$. Before this is done,
equation (\ref{expan1}) is simplified by several means.

With the Nabla operator in spherical coordinates

\bq
\nabla
=\er\parr + \frac{1}{r}\,\etheta\parth + \frac{1}{r\sin\vartheta}\,\ephi\parphi
\eq

where $\partial_j\equiv{\partial}/{\partial j}$, the velocity components read:

\bq
v_r = \mysum \flm \ylm  \label{expan_vr}
\eq
\bqa
\left(
\eqalign{\vth \\ \vphi}
\right)
=\mysum \left\{ \glm
\left(
\eqalign{\parth\ylm \\ \frac{\ii m}{\sth}\ylm }
\right) + \hlm
\left(
\eqalign{-\frac{\ii m}{\sth}\ylm \\ \parth\ylm }
\right)  \right\}       \label{expan_vthphi}
\eqa

With (\ref{ylm_minus}) and the fact that the velocity field is real valued
it follows

\bq
\glmmi = (-1)^m\glm,\quad \hlmmi = (-1)^m\hlm \,. \label{gh_minus}
\eq

Furthermore, when the symmetry of the problem with respect to the $xz$-plane, \emph{i.e.}

\bq
\vth(-\varphi) = \vth(\varphi),\qquad   \vphi(-\varphi) = -\vphi(\varphi)\label{symm1}
\eq

is taken into account, it can be shown with the aid of relations (\ref{ylm_minus})
and (\ref{gh_minus}) that

\bqa\fl
\left(\eqalign{\vth \\ \vphi }\right)
=\sum_{\ell=1}^\infty\sum_{m=0}^{\ell}\!\strut^{'} 2\klm\!\left\{
\, \glm  \left( \eqalign{\cos(m\varphi)\parth\plm\\ \sin(m\varphi)\frac{-m}{\sth}\plm}
                \right)
- \hlm \left(
                \eqalign{\cos(m\varphi)\frac{-m}{\sth}\plm\\ \sin(m\varphi)\parth\plm}
                \right) \right\}
\nonumber\\
\equiv 2\mysum \Big\{ \glm\vAlm+\hlm\vBlm \Big\}\,.  \label{expan_real}
\eqa

The prime at the second sum indicates that the terms with $m=0$ are divided by two.

When the boundary conditions are applied
it will be important that the two velocity components of (\ref{expan_real})
always be considered together, because $\vAlm=\vAtp$ and $\vBlm=\vBtp$ fulfil the
orthogonality relations

\bqa
\la\vAlm,\vBlmp\ra &=& 0                                \label{orth2}   \\
\la\vAlm,\vAlmp\ra &=& \la\vBlm,\vBlmp\ra = \frac{1}{2}\ell(\ell+1)\delta_{\ell\ell'}
\delta_{mm'}               \label{orth3}
\eqa

with the vector inner product

\bq
\la\bi{X}_1,\bi{X}_2\ra
:= \int_0^{2\pi}\dphi\int_0^\pi\dth\,\sth\,(\bi{X}_1^*)\transp\bi{X}_2\,,
\label{vecprod}
\eq

where $^*$ denotes the complex conjugate and $\transp$ the transpose of the vector.
By computing the inner product of $\bi{A}_{\ell'm'}$ or $\bi{B}_{\ell'm'}$ with
(\ref{expan_real}) one can reduce the infinite series to one function $g_{\ell'm'}(r)$
or $h_{\ell'm'}(r)$, respectively.
If $\vth$ and $\vphi$ were considered seperately, it would not be possible to get at
the radial functions, because $\parth P_{\ell'm'}$ and $\pm\frac{\ii m}{\sth}\plm$ alone
are \emph{not} orthogonal.

Now putting the expansions (\ref{expan_vr}) and (\ref{expan_real}) into the basic
equations (\ref{cont}) and (\ref{stokesrot}) gives the following ordinary differential
equations for the radial functions with $\ell>0$ ($g_{00}(R)=h_{00}(R)\equiv 0$ can be
assumed w.l.o.g.):

\bq
f_{00}' + \frac{2}{r} f_{00} = 0                \label{difffnull}
\eq
\bqa\fl
\frac{r}{\ell(\ell+1)}\flm'''' + \frac{8}{\ell(\ell+1)}\flm'''
+\frac{2}{r} \left[ \frac{6}{\ell(\ell+1)} - 1 \right]\flm''
- \frac{4}{r^2}\flm'  +\frac{1}{r^3} \Bigl[ \ell(\ell+1) - 2 \Bigr]\flm \nonumber\\= 0
\label{difff}
\eqa

\bq
\glm(r) = \frac{1}{\ell(\ell+1)} \Bigl[ r\flm' + 2\flm \Bigr]  \label{diffg}
\eq

\bq
\hlm'' + \frac{2}{r}\hlm' - \frac{\ell(\ell+1)}{r^2}\hlm = 0   \label{diffh}
\eq

These equations are solved by a power law ansatz which together with the requirement
that the velocity be finite as $r\to\infty$ leads to

\bqa
h_{\ell m}(r) &= \frac{a_{\ell m}}{r^{\ell+1}},\qquad \ell>0   \label{hlm_a}\\
f_{00}(r) &= \frac{d_{00}}{r^2}   \\
f_{1m}(r) &= b_{1m}+\frac{c_{1m}}{r}+\frac{d_{1m}}{r^3} \\
g_{1m}(r) &= b_{1m}+\frac{c_{1m}}{2r}-\frac{d_{1m}}{2r^3}       \label{g1m_cd}
\eqa

and for $\ell>1$:

\bqa
f_{\ell m}(r)
  &=& \frac{c_{\ell m}}{r^\ell}+\frac{d_{\ell m}}{r^{\ell+2}}\\
g_{\ell m}(r) &=& \frac{-1}{\ell(\ell+1)}\left[\frac{(\ell-2)c_{\ell m}}{r^\ell}
+\frac{\ell d_{\ell m}}{r^{\ell+2}}\right]      \label{glm_cd}
\eqa

The coefficients $\alm$, $b_{1m}$, $\clm$, and $\ddlm$ are determined by successively
applying the remaining boundary conditions. In the following section, the ferrofluid
drop will be treated as a solid sphere. Its angular velocity $\itom$ is introduced as
a parameter that will have to be determined by the equality of viscous and magnetic
torques generated by external field and surrounding liquid.

It should be noted here, that the orthogonality relations (\ref{orth2}) and (\ref{orth3})
would not be valid if the $\vartheta$-integral within the scalar product (\ref{vecprod})
were only carried out up to $\vartheta=\pi/2$. On the other hand, the liquid only
occupies the lower half-space in the given problem, so we perform a little trick in
order to be able to integrate over the whole sphere, \emph{i.e.}, we take advantage of our
equations being linear and employ the superposition principle by adding
the {\it mirror image} of our problem with respect to the $xy$-plane (fluid above, void
below the sphere). The problem can be solved in this way and the resulting flow field in
the upper half space is simply neglected in the end.

Within the framework of this "mirror image construction" the following boundary
conditions are employed:

\begin{itemize}
\item
Navier slip at the sphere surface
\bq
\left[\parr-\frac{1}{R}\right]\left.
\left(
\eqalign{\vth\\ \vphi}
\right)\right|_{r=R}
= \frac{1}{\ls}\left[\left(
\eqalign{\vth(r=R)\\ \vphi(r=R)}
\right)-\bi{U}\right]   \label{BCrot}
\eq

with the slip length $\ls\ll R$ and the velocity $\bi{U}$ of the sphere surface

\bq
\bi{U}=\left\{
\eqalign{
\phantom{-R\vecom} 0 & \qquad \textrm{for }\vartheta= {\pi}/{2}   \\
\phantom{-}R\vecom\times \er & \qquad \textrm{for }\vartheta< {\pi}/{2} \\
-R\vecom\times \er & \qquad \textrm{for }\vartheta> {\pi}/{2} }
\right.
\eq

implying

\bq
v_r(r=R) = 0\,.
\eq

\item
Flat "interface":

\bq
\vth\left(\vartheta=\frac{\pi}{2}\right) = 0
\qquad \forall \,r\geq R  \label{BCcontsymm}
\eq

\item
No resulting (viscous) force on the sphere:

\bqa
F_i &=& \int_0^{2\pi}\dphi\int_0^{\pi/2} \dth\sth
\sum_j\sigma_{rj}(r=R)\,\ej\cdot\ei = 0         \label{BCforce}
\eqa

with $i\in\{x,y,z\}$, $j\in\{r,\vartheta,\varphi\}$, and $\ei$, $\ej$ the unit vectors in
respective direction. The relevant components of the viscous stress tensor $\sigma_{rj}$
are taken as defined in \cite{LLhydro}.

As is obvious from the given symmetry, only $F_x$ will be different from zero and
thereby determine the last coefficient.

Since the magnetic field only creates a torque but no linear force, this
boundary condition provides the requirement of unaccelerated translational motion.

\end{itemize}

%%%%%%%%%%%%%%%%%%%%%%%%%%%%%%%%%%%%%%%%%%%%%%%%%%%%%%%%%%%%%%%%%%%%%%%%%%%%%%%%%%%%%%%
\subsection{Applying the boundary conditions}{bcapp}

The first coefficients are determined by the $r$-component of the Navier slip condition

\bq
v_r(R) = \mysum\flm(R)\,\ytp = 0
\eq

and the orthogonality of the scalar spherical harmonics $\ylm$ \cite{AbSt}:

\bqa
\flm(R) &=& 0\quad\forall\ell,m         \\
&\Rightarrow&
\eqalign{d_{00\phantom{,}} = 0          \\
d_{1m} = -R^3b_{1m}-R^2c_{1m}   \\
d_{\ell m} = -R^2c_{\ell m},\quad \ell>1}
\eqa

The coefficients $c_{\ell m}$ and $a_{\ell m}$ are obtained by applying
the appropriate vector inner product to the $\vartheta$- and $\varphi$-component of
the Navier slip condition

\bqa\fl
\left[ 1+\frac{\ls}{R}-\ls\parr \right]
\sum_{\ell=1}^\infty\sum_{m=0}^{\ell}\!\strut^{'} 2\klm\!\!
\left.\left[
\glm  \left(    \eqalign{\cos(m\varphi)\parth\\ \sin(m\varphi)\frac{-m}{\sth}}
                        \right)
- \hlm \left(           \eqalign{ \cos(m\varphi)\frac{-m}{\sth}\\ \sin(m\varphi)\parth}
                        \right) \right]\!\plm\,\right|_R
        \nonumber\\
= \left\{
\eqalign{
\phantom{-R\itom(\cph\,\etheta)} 0 & \qquad \textrm{for } \vartheta= {\pi}/{2}\\
\phantom{-}R\itom(\cph\,\etheta-\cth\sph\,\ephi) & \qquad  \textrm{for }
\vartheta< {\pi}/{2}   \\
-R\itom(\cph\,\etheta-\cth\sph\,\ephi) & \qquad  \textrm{for } \vartheta> {\pi}/{2}}
\right.
\eqa

which is done here exemplary for the scalar product with $\vAlmp(\vartheta,\varphi)$
as defined in \ref{vecprod}. The orthogonalities of the sine and cosine
functions yield

\bq
\left( 1+\frac{\ls}{R} \right)\glm(R) - \ls \,\glm'(R) = 0 \qquad \forall\;m\neq\pm1
\eq

and

\bqa\fl
\left[\left( 1+\frac{\ls}{R} \right)\glm(R) - \ls \,\glm'(R) \right] \ell(\ell+1) =
     \nonumber\\
\pi\itom R\kleins\int_0^{\pi/2}\dth\sth
   \Bigl[\parth + \cot\vartheta\Bigr]\pleins(\cth)      \nonumber\\
-\pi\itom R\kleins\int_{\pi/2}^{\pi}\dth\sth
   \Bigl[\parth + \cot\vartheta\Bigr]\pleins(\cth).
\eqa

Now from \cite{AbSt} one finds

\bq
\parth\pleins + \cot\vartheta\pleins = \ell(\ell+1)\plnull
\eq

and

\bqa
\int_0^1\du P_{\ell}(u) = \phantom{-}\int_{-1}^0\du P_{\ell}(u),&\qquad
\ell\textrm{ even}\\
\int_0^1\du P_{\ell}(u) = -\int_{-1}^0\du P_{\ell}(u)= (-1)^\frac{\ell-1}{2}
\frac{(\ell-2)!!}{(\ell+1)!!}
,&\qquad \ell\textrm{ odd}
\eqa

so that with the definition of $K_{\ell m}$ according to (\ref{klmdef}) one obtains

\bq
\left[\left( 1+\frac{\ls}{R} \right)g_{\ell 1}(R) - \ls \,g_{\ell 1}'(R) \right]= 0
\qquad \forall\;\ell\textrm{ even}
\eq

and for odd $\ell$

\bq\fl
\left[\left( 1+\frac{\ls}{R} \right)g_{\ell 1}(R) - \ls \,g_{\ell 1}'(R) \right]=
\itom R \,\sqrt{\frac{(2\ell+1)\pi}{\ell(\ell+1)}}\, (-1)^{\frac{\ell+1}{2}}
\,\frac{(\ell-2)!!}{(\ell+1)!!}\,.
\eq

With (\ref{g1m_cd}) and (\ref{glm_cd}) this gives in detail

\bqa
b_{10} \left[ \frac{3}{2}+3\frac{\ls}{R} \right] + \frac{c_{10}}{R}
\left[ 1+3\frac{\ls}{R} \right] = 0 \\
b_{1,\pm 1} \left[ \frac{3}{2}+3\frac{\ls}{R} \right] + \frac{c_{1,\pm 1}}{R}
\left[ 1+3\frac{\ls}{R} \right]
= \mp \itom R\, \sqrt{\frac{3\pi}{2}}
\eqa
\bq\eqalign{
c_{\ell m} = 0  \;\qquad\forall \;m\neq \pm 1   \\
c_{\ell, \pm 1} = 0  \qquad\forall \;\ell \textrm{ even}        \\
c_{\ell,\pm 1} = \pm\frac{\itom}{2} \,\sqrt{\pi\ell(\ell+1)(2\ell+1)}\,
\frac{R^{\ell+1}\,(-1)^\frac{\ell+1}{2}}{1+(2\ell+1)\frac{\ls}{R}}\cdot
\frac{(\ell-2)!!}{(\ell+1)!!},\quad\ell\textrm{ odd}.}
\eq

The condition of a flat "interface" reads

\bqa\fl
g_{10}(r)K_{10}\Bigl[\parth P_{10}(\cth)\Bigr]_{\vartheta=\frac{\pi}{2}}
+ \sum_{m=\pm1} \sum_{\stackrel{\ell=1}{\ell\textrm{ \scriptsize odd}}}^\infty
\glm(r)\klm\Bigl[\parth\plm(\cth)\Bigr]_{\vartheta=\frac{\pi}{2}}    \nonumber\\
+ \sum_{m=\pm1}\sum_{\stackrel{\ell=2}{\ell\textrm{ \scriptsize even}}}^\infty
m\hlm(r)\klm\plm(0)=0 .
\eqa

The sums vanish completely due to properties of the Legendre functions at zero
\cite{AbSt}, so that only the first term remains, giving

\bq
b_{10}\left[ 1+\frac{1}{2}\frac{R^3}{r^3} \right] + \frac{c_{10}}{2r}
\left[ 1+\frac{R^2}{r^2} \right] = 0\,.
\eq

Since this equation must be valid for arbitrary $r$ it follows $b_{10}=0=c_{10}$.\\
In order to evaluate the force condition

\bqa\fl
F_x = R^2 \!\int_0^{2\pi}\dphi\int_0^{\pi/2}\dth\sth
\Bigl[ \sigrr(R)\sth\cph  \nonumber\\
+\sigrt(R)\cth\cph  -\sigrp(R)\sph  \Bigr]
= 0
\eqa

the following integrals are needed:

\bqa\fl
\int_0^{\pi/2}\dth\,\sth\left[\frac{1}{\sth}  +\cth\parth\right] \pleins(\cth)
= \int_0^{\pi/2}\dth\sqth\pleins(\cth)   = \frac{4}{3}\,\delta_{\ell 1}
\eqa

Then the last coefficients are given by

\bq
b_{1,\pm 1} = \mp\,\frac{\itom R}{1+2\frac{\ls}{R}}\,\sqrt{\frac{\pi}{6}}\,.
\eq
%%%%%%%%%%%%%%%%%%%%%%%%%%%%%%%%%%%%%%%%%%%%%%%%%%%%%%%%%%%%%%%%%%%%%%%%%%%%%%%%%%%%%%%

\section{Resulting flow fields for the liquid half-sphere model}\label{thirdapp}

\bqa\fl
v_r^{(\mathrm{i})}
= \frac{3}{4}
    \frac{\mathfrak{M} R}{2\eta^{(\mathrm{o})}+3\eta^{(\mathrm{i})}}
    \, \sth\,\cph        \left[ \frac{r^2}{R^2}-1 \right] \nonumber\\
+\frac{\mathfrak{M} R\cph}{\eta^{(\mathrm{o})}+\eta^{(\mathrm{i})}}
    \sum_{\stackrel{\ell=3}{\ell\textrm{ \scriptsize odd}}}^\infty
    \pleins(\cth)\,
    \frac{r^{\ell-1}}{R^{\ell-1}}\, (-1)^\frac{\ell-1}{2}
    \left[ \frac{r^2}{R^2}-1 \right]
    \frac{(\ell-2)!!}{(\ell+1)!!}       %\label{vr_i}
\eqa

\bqa\fl
v_r^{(\mathrm{o})}
= \frac{1}{2}
    \frac{\mathfrak{M} R}{2\eta^{(\mathrm{o})}+3\eta^{(\mathrm{i})}}
    \,\sth\cph\left[ 1-\frac{R^3}{r^3} \right]  \nonumber\\
+\frac{\mathfrak{M} R\cph}{\eta^{(\mathrm{o})}+\eta^{(\mathrm{i})}}
    \sum_{\stackrel{\ell=3}{\ell\textrm{ \scriptsize odd}}}^\infty
    \pleins(\cth)\,   \frac{R^\ell}{r^\ell}
    \left[ 1-\frac{R^2}{r^2} \right](-1)^\frac{\ell-1}{2}\,
    \frac{(\ell-2)!!}{(\ell+1)!!}
\eqa

\bqa\fl
\left(
\eqalign{\vth^{(\mathrm{i})} \\ \vphi^{(\mathrm{i})} }
\right)
= \frac{3}{4}
    \frac{\mathfrak{M} R}{2\eta^{(\mathrm{o})}+3\eta^{(\mathrm{i})}}
    \left(
    \eqalign{ \cph\cth\\ -\sph}
    \right)
    \left[ 2\frac{r^2}{R^2}-1 \right]   \nonumber\\
+ \frac{\mathfrak{M} R}{\eta^{(\mathrm{o})}+\eta^{(\mathrm{i})}}
    \left(
    \eqalign{ \cph\,\parth  \\ -\sph/\sth}
    \right)     \nonumber\\
    \times\sum_{\stackrel{\ell=3}{\ell\textrm{ \scriptsize odd}}}^\infty
    \pleins(\cth) \,
    \frac{r^{\ell-1}}{R^{\ell-1}}
    \left[ \frac{(\ell+3)}{(\ell+1)} \frac{r^2}{R^2}-1 \right]
    \frac{(-1)^\frac{\ell-1}{2}}{\ell}\,
    \frac{(\ell-2)!!}{(\ell+1)!!}       \nonumber\\
+ 2\mathfrak{M} R   \left(
    \eqalign{ -\cph/\sth  \\  \sph \,\parth}
    \right)     \nonumber\\
    \times    \sum_{\stackrel{\ell=2}{\ell\textrm{ \scriptsize even}}}^\infty
    \frac{\pleins(\cth)\,(-1)^\frac{\ell}{2}}{(\ell+2)\eta^{(\mathrm{o})}+(\ell-1)%
    \eta^{(\mathrm{i})}}\,
    \frac{r^\ell}{R^\ell}\,
    \frac{(2\ell+1)(\ell-3)!!}{\ell(\ell+1)(\ell+2)!!}
\eqa

\bqa\fl
\left(
\eqalign{\vth^{(\mathrm{o})} \\ \vphi^{(\mathrm{o})}}
\right)
= \frac{1}{2}
    \frac{\mathfrak{M} R}{2\eta^{(\mathrm{o})}+3\eta^{(\mathrm{i})}}
    \left(
    \eqalign{ \cph\cth \\ -\sph}
    \right)
    \left[ 1+\frac{1}{2}\frac{R^3}{r^3} \right]         \nonumber\\
+\frac{\mathfrak{M} R}{\eta^{(\mathrm{o})}+\eta^{(\mathrm{i})}}
    \left(
    \eqalign{ \cph \,\parth \\-\sph/\sth}
    \right)     \nonumber\\
    \times    \sum_{\stackrel{\ell=3}{\ell\textrm{ \scriptsize odd}}}^\infty
    \pleins(\cth)\,
    \frac{R^\ell}{r^\ell}
    \left[ (2-\ell)+\ell\frac{R^2}{r^2}\right]
    \frac{(-1)^\frac{\ell-1}{2}}{\ell(\ell+1)}\,
    \frac{(\ell-2)!!}{(\ell+1)!!}       \nonumber\\
+ 2\mathfrak{M} R
    \left(
    \eqalign{  -\cph/\sth \\ \sph \,\parth}
    \right)     \nonumber\\
    \times    \sum_{\stackrel{\ell=2}{\ell\textrm{ \scriptsize even}}}^\infty
    \frac{\pleins(\cth)\,(-1)^\frac{\ell}{2}}%
        {(\ell+2)\eta^{(\mathrm{o})}+(\ell-1)\eta^{(\mathrm{i})}}\,
    \frac{R^{\ell+1}}{r^{\ell+1}}\,
    \frac{(2\ell+1)(\ell-3)!!}{\ell(\ell+1)(\ell+2)!!}
\eqa

\end{appendix}

%%%%%%%%%%%%%%%%%%%%%%%%%%%%%%%%%%%%%%%%%%%%%%%%%%%%%%%%%%%%%%%%%%%%%%%%%%%%%%%%%%%%%%%
%%%%%%%%%%%%%%%%%%%%%%%%%%%%%%%%%%%%%%%%%%%%%%%%%%%%%%%%%%%%%%%%%%%%%%%%%%%%%%%%%%%%%%%

\end{document}

%%%%%%%%%%%%%%%%%%%%%%%%%%%%%%%%%%%%%%%%%%%%%%%%%%%%%%%%%%%%%%%%%%%%%%%%%%%%%%%%%%%%%%%
%%%%%%%%%%%%%%%%%%%%%%%%%%%%%%%%%%%%%%%%%%%%%%%%%%%%%%%%%%%%%%%%%%%%%%%%%%%%%%%%%%%%%%%

%bibliography Reinhard
%\section*{References}
%\def \bib#1{C:/RR/texte/paper/bib/#1}
%\bibliographystyle{\bib{RR_prsty}} %apsrev  unsrt
%\bibliography{\bib{xr},\bib{mfcom},\bib{mfrr},\bib{drop_lit}}

\bibitem{rosensweig1985}
R.~E. Rosensweig, {\em Ferrohydrodynamics} (Cambridge University Press,
  Cambridge, 1985).

\bibitem{bacri1994}
J.~C. Bacri, A. Cebers, S. Lacis, and R. Perzynski, Phys. Rev. Lett. {\bf 72},
  2705  (1994).

\bibitem{cebers1995}
A. Cebers, S. Lacis, and J. Brazilian, J. Phys. {\bf 25},  101  (1995).

\bibitem{lebedev1997}
K. Morozov and A.~V. Lebedev, Sov. Phys. JETP {\bf 65},  160  (1997).

\bibitem{morozov1997}
K.~I. Morozov, JETP {\bf 85},  728  (2000).

\bibitem{sandre1999}
O. Sandre, J. Broways, R. Perzynski, J.-C. Bacri, V. Cabuil, and R.~E.
  Rosensweig, Phys. Rev. E {\bf 59},  1736  (1999).

\bibitem{cebers2002}
A. Cebers, Phys. Rev. E {\bf 66},  061402  (2002).

\bibitem{morozov2000}
K. Morozov and A.~V. Lebedev, JETP {\bf 91},  1029  (2000).

\bibitem{lebedev2003}
A.~V. Lebedev, A. Engel, K.~I. Morozov, and H. Bauke, New J. Phys. {\bf 5},
  57.1  (2003).

\bibitem{blums1997}
E. Blums, A. Cebers, and M. Maiorov, {\em Magnetic Fluids} (Walter de Gruyter
  and Co., Berlin, New York, 1997).

\bibitem{krauss2005}
R. Krauss, M. Liu, B. Reimann, R. Richter, and I. Rehberg, Appl. Phys. Lett.
  {\bf 86},  024102  (2005).

\bibitem{krauss2006}
R. Krauss, M. Liu, B. Reimann, R. Richter, and I. Rehberg, N. J. Phys. {\bf 8},
   18  (2006).

\bibitem{boennemann2003b}
H. B\"onnemann, W. Brijoux, R. Brinkmann, N. Matoussevitch, and N. Wald\"ofner,
  German patent DE 10227779.6., 2003.

\bibitem{huh_scri}
C. Huh and L.~E. Scriven, Journal of Colloid and Interface Science {\bf 35},
  85  (1971).

\bibitem{deGen1}
P.~G. de~Gennes, Colloid and Polymer Science {\bf 264},  463  (1986).

\bibitem{navier}
C.-L. Navier, M\'emoires de l'{A}cad\'emie des {S}ciences de l'{I}nstitut de
  {F}rance {\bf 6},  389  (1823).

\bibitem{dussan1}
E.~B. Dussan~V., Journal of Fluid Mechanics {\bf 77},  665  (1976).

\bibitem{pis_rub}
L.~M. Pismen and B.~Y. Rubinstein, Langmuir {\bf 17},  5265  (2001).

\bibitem{huh_mas}
C. Huh and S.~G. Mason, Journal of Fluid Mechanics {\bf 81},  401  (1977).

\bibitem{hocking2}
L.~M. Hocking, Journal of Fluid Mechanics {\bf 79},  209  (1977).

\bibitem{cox}
R.~G. Cox, Journal of Fluid Mechanics {\bf 168},  169  (1986).

\bibitem{oneill}
M.~E. O'Neill, K.~B. Ranger, and H. Brenner, Physics of Fluids {\bf 29},  913
  (1986).

\bibitem{denn}
M.~M. Denn, Annual Review of Fluid Mechanics {\bf 33},  265  (2001).

\bibitem{jos_tab}
P. Joseph and P. Tabeling, Physical Review E {\bf 71},  035303  (2005).

\bibitem{schmatko}
T. Schmatko, H. Hervet, and L. Leger, Physical Review Letters {\bf 95},  244501
   (2005).

\bibitem{fetzer_etal}
R. Fetzer, M. Rauscher, A. M\"unch, B. Wagner, and K. Jacobs, Europhysics
  Letters {\bf 4},  638  (2006).

\bibitem{MoFe}
P.~M. Morse and H. Feshbach, {\em Methods of theoretical physics} (McGraw-Hill
  book company, Inc., New York, 1953), Vol.~2, chapter 13.

\bibitem{cohentann}
C. Cohen-Tannoudji, B. Diu, and F. Lalo\"e, {\em Quantenmechanik}, 2 ed. (de
  Gruyter, Berlin, 1999).

\bibitem{LLhydro}
L.~D. Landau and E.~M. Lifschitz, {\em Lehrbuch der {T}heoretischen {P}hysik},
  5 ed. (Akademie Verlag, Berlin, 1991), Vol.~6, hydrodynamik.

\bibitem{LLmag}
L.~D. Landau and E.~M. Lifschitz, {\em Lehrbuch der {T}heoretischen {P}hysik},
  5 ed. (Akademie Verlag, Berlin, 1990), Vol.~8, elektrodynamik der Kontinua.

\bibitem{shliomis}
M.~I. Shliomis, Soviet Physics-Uspekhi {\bf 17},  153  (1974).

\bibitem{pamme2006}
N. Pamme, Lab Chip {\bf 6},  24  (2006).

\bibitem{nguyen2006}
N.-T. Nguyen, K.~M. Ng, and X. Huang, Appl. Phys. Lett. {\bf 89},  052509
  (2006).

\bibitem{thorsen2001}
T. Thorsen, R.~W. Roberts, F.~H. Arnold, and S.~R. Quake, Phys. Rev. Lett. {\bf
  86},  4163  (2001).

\bibitem{pekas2004}
N. Pekas, M.~D. Porter, M. Tondra, A. Popple, and A. Jander, Appl. Phys. Lett.
  {\bf 85},  4783  (2004), von Robert.

\bibitem{AbSt}
M. Abramowitz and I.~A. Stegun, {\em Handbook of mathematical functions}, 14
  ed. (Dover, New York, 1970).